\documentclass{article}
\usepackage{amssymb}

\begin{document}

\title{Nuclear deformation and the two neutrino double-$\beta $ decay in $%
^{124,126} $Xe,$^{128,130}$Te, $^{130,132}$Ba and $^{150}$Nd isotopes}
\author{S. Singh$^{1}$, R. Chandra$^{1}$, P. K. Rath$^{1}$, P. K. Raina$^{2}$ and J.
G. Hirsch$^{3}$ \\
$^{1}$Department of Physics, University of Lucknow, Lucknow-226007, India\\
$^{2}$Department of Physics and Meteorology, IIT, Kharagpur-721302, India\\
$^{3}$Instituto de Ciencias Nucleares, Universidad Nacional Aut\'{o}noma de
M\'{e}xico,\\
A.P. 70-543, M\'{e}xico 04510 D.F., M\'{e}xico}
\date{}
\maketitle

\begin{abstract}
The two neutrino double beta decay of $^{124,126}$Xe,$^{128,130}$Te, $%
^{130,132}$Ba and $^{150}$Nd isotopes is studied in the Projected
Hartree-Fock-Bogoliubov (PHFB) model. Theoretical 2$\nu $ $\beta ^{-}\beta
^{-}$ half-lives of $^{128,130}$Te, and $^{150}$Nd isotopes, and 2$\nu $ $%
\beta ^{+}\beta ^{+}$, 2$\nu $ $\beta ^{+}EC$ and 2$\nu $ $ECEC$ for $%
^{124,126}$Xe and $^{130,132}$Ba nuclei are presented. Calculated
quadrupolar transition probabilities B(E2: $0^+\rightarrow 2^+$), static
quadrupole moments and $g$ factors in the parent and daughter nuclei
reproduce the experimental information, validating the reliability of the
model wave functions. The anticorrelation between nuclear deformation and
the nuclear transition matrix element $M_{2\nu} $ is confirmed.

PACS Numbers: 23.40.Hc, 21.60.Jz, 23.20.-g, 27.60.+j
\end{abstract}

\section{Introduction}

The nuclear double beta decay ($\beta \beta $) is a weak process in which
two neutrons (protons) inside a nucleus decay into two protons (neutrons),
emitting two electrons (positrons) \cite{pri52,bri78}. In the
lepton-number-conserving process, namely two neutrino double beta (2$\nu $ $%
\beta \beta $) decay two antineutrinos (neutrinos) are also emitted [3-8].
It has been observed in several nuclei [10-13]. Given that lepton number is
not a gauge symmetry, several extensions of the standard model predict its
violation. In this neutrinoless double beta (0$\nu $ $\beta \beta $) decay
no neutrinos are emitted. This 0$\nu $ $\beta \beta $ decay has never been
observed, although controversial evidence has been published \cite{kla04}.
If positively detected, it would provide unique information about the
Majorana mass of the neutrino, and for the absolute scale of neutrino masses
[15-20].

Theoretical calculations of $\beta \beta $ decay represent one of the
hardest challenges ever faced in nuclear physics. The process is
definitively non-collective, and proceed through strongly suppressed
channels, which are very sensitive to details of the wave functions of the
parent and daughter nuclei. While specific models have been built which are
able to adjust (postdict) most of the observed 2$\nu $ $\beta \beta $ decay
half lives, they usually have parameters which allow different results as
well, making their predictions unreliable. A case by case analysis, studying
each nuclei with the best available model, and describing as much nuclear
observables as possible within the same theoretical scheme, seems to be a
sensible approach [21-23].

The 2$\nu $ $\beta \beta $ decay can occur in four different modes, namely
double electron ($\beta ^{-}\beta ^{-}$) emission, double positron $(\beta
^{+}\beta ^{+})$ emission, electron-positron conversion $(\beta ^{+}EC)$ and
double electron capture $(ECEC)$. The later three processes are
energetically competing and we shall refer to them all generically as
positron double beta decay (e$^{+}$DBD) modes. Nuclear matrix elements
(NMEs) $M_{2\nu }$ associated with the 2$\nu $ $\beta \beta $ decay can be
extracted directly from the observed half-lives of eleven nuclei undergoing 2%
$\nu \beta ^{-}\beta ^{-}$ decay out of 35 possible candidates \cite{tre95}.
In case of 2$\nu $ e$^{+}$DBD modes, experimental limits on half-lives have
already been given for 24 out of 34 possible isotopes \cite{tre95}.

In all $\beta \beta $ decay emitters, which are even-\textit{Z} and even-%
\textit{N} nuclei, the pairing of like nucleons plays a fundamental role,
energetically inhibiting the decay to the intermediate odd-odd nuclei. The
quadrupole-quadrupole interaction drives nuclei to deformed shapes when both
proton and neutron shells are open. The interplay between this two major
components of the effective nuclear interaction can generate quite complex
energy spectra. In the mass region \textit{A}$\approx $130 it can be seen
that, while Te isotopes have a vibrational excitation spectra, Xe and Ba
isotopes develop rotational bands. %
%
The mass region $A\thicksim 150$ offers an example of shape transition i.e.
the sudden onset of deformation at neutron number $N$=90. Nuclei range from
spherical to well deformed, with large static quadrupole moments. %
%
Thus, it is expected that pairing and deformation degrees of freedom will
play some crucial role in the structure of $^{124,126,128,130,132}$Xe, $%
^{124,126,128,130}$Te, $^{130,132}$Ba, $^{150}$Nd and $^{150}$Sm nuclei. %
%
The $\beta \beta $ decay can be studied in the same framework as many other
nuclear properties and decays. A vast amount of data has been collected over
the past years concerning the level energies as well as electromagnetic
properties through experimental studies involving in-beam $\gamma $-ray
spectroscopy. The availability of data permits a rigorous and detailed
critique of the ingredients of the microscopic model that seeks to provide a
description of nuclear $\beta \beta $ decay. %
%

Theoretical studies predict that deformation plays a crucial role in case of
2$\nu $ $\beta ^{-}\beta ^{-}$ decay of $^{100}$Mo and $^{150}$Nd \cite
{gri92,suh94}. Auerbach \textit{et al.} \cite{aue93} and Troltenier \textit{%
et al.} \cite{tro96} have already shown that there exists an inverse
correlation between the Gamow-Teller (GT) strength and quadrupole moment.
The effect of deformation on the distribution of the Gamow-Teller and $\beta 
$-decay properties has been studied using a quasiparticle Tamm-Dancoff
approximation (TDA) based on deformed Hartree-Fock (DHF) calculations with
Skyrme interactions \cite{fri95}, a deformed self consistent HF+RPA method
with Skyrme type interactions \cite{sar98}. The comparison of the
experimental GT strength distribution $B(GT)$ from its decay with the
results of QRPA calculations was employed as a novel method of deducing the
deformation of the $N=Z$ nucleus $^{76}$Sr \cite{nach04}. The effect of
deformation on the 2$\nu $ $\beta ^{-}\beta ^{-}$ decay for ground-state
transition $^{76}$Ge $\rightarrow $ $^{76}$Se was studied in the framework
of the deformed QRPA with separable GT residual interaction \cite{pac04}. A
deformed QRPA formalism to describe simultaneously the energy distributions
of the single beta GT strength and the 2$\nu $ $\beta \beta $ decay matrix
elements, using deformed Woods-Saxon potentials and deformed Skyrme
Hartree-Fock mean fields was developed \cite{alv04}.

In all these works calculations are performed in the intrinsic basis, where
angular momentum is not a good quantum number. The Projected
Hartree-Fock-Bogoliubov (PHFB) model offers, in this sense, a competitive
alternative. On one hand, the PHFB model restores the rotational symmetry,
providing very reliable wave functions for the parent and daughter $\beta
\beta $ decay emitters. On the other hand, in its present version the PHFB
model is unable to provide information about the structure of the
intermediate odd-odd nuclei, and, in particular, on the single $\beta $
decay rates and the distribution of Gamow-Teller strength. Notwithstanding,
the PHFB model has been successfully applied to the 2$\nu $ $\beta ^{-}\beta
^{-}$ decay of many emitters in the mass region $A\thicksim 100$, where it
was possible to describe, in the same context, the lowest excited states of
the parent and daughter nuclei, as well as their electromagnetic transition
strengths, and to reproduce their measured $\beta \beta $ decay rates on the
other \cite{cha05}.

A quantitative description of the 2$\nu $ $\beta ^{-}\beta ^{-}$ decay \cite
{cha05} and 2$\nu $ e$^{+} $DBD modes \cite{rai06} of nuclei in the mass
region $A\sim $100 for the $0^{+}\rightarrow 0^{+}$ transition has been
obtained. The same formalism allowed the analysis of other observed nuclear
properties, including the yrast spectra, the reduced transition
probabilities $B(E2$: $0^{+}\rightarrow 2^{+})$, the static quadrupole
moments $Q(2^{+})$ and the $g$ -factors $g(2^{+})$ of both parent and
daughter nuclei. The study was performed using the PHFB model. Its
application, in conjunction with the summation method, has motivated the
present study of the 2$\nu $ $\beta ^{-}\beta ^{-}$ decay of $^{128,130}$
Te, $^{150}$Nd and 2$\nu $ e$^{+}$DBD modes of $^{124,126}$Xe and $%
^{130,132} $Ba isotopes. The 2$\nu $ e$^{+}$DBD modes of $^{124,126}$Xe and $%
^{130,132} $Ba isotopes were studied in the earlier work of Shukla \textit{%
et al.} \cite{shuk07}. In the present work, the HFB wave functions are
generated with higher accuracy and in these nuclei, it is observed that the
results are very much dependent on the later.

By the use of the pairing plus quadrupole-quadrupole \textit{PPQQ}
interaction \cite{bar68} %
%
the interplay between sphericity and deformation can be studied. In this
way, the PHFB formalism, employed in conjunction with the \textit{PPQQ}
interaction, is a convenient choice to examine the explicit role of
deformation on the NTME\ $M_{2\nu }$. A strong dependence of the $M_{2\nu }$
on the quadrupole deformation was found varying the strength of the
quadrupole-quadrupole interaction, for the 2$\nu $ $\beta ^{-}\beta ^{-}$
decay \cite{cha05} and 2$\nu $ e$^{+}$DBD modes \cite{rai06} of nuclei in
the mass region $A\sim $100. %
%
The anticorrelation between the quadrupolar deformation and the $\beta \beta 
$ decay transition amplitude is analyzed in detail.

The present paper is organized as follows. The theoretical formalism to
calculate the half-lives of 2$\nu $ $\beta \beta $ modes has been given in a
number of reviews \cite{doi92, suh98}. Details of the PHFB study were
presented in previous publications, for 2$\nu $ $\beta ^{-}\beta ^{-}$ decay 
\cite{cha05} and 2$\nu $ e$^{+}$DBD modes \cite{rai06} of nuclei in the mass
region $A\sim $100. Details of the mathematical expressions used to
calculate the spectroscopic properties of nuclei in the PHFB model have been
given by Dixit \textit{et al. }\cite{dix02}. Here, we briefly outline the
relevant results in Section 2. In Section 3, we present a detailed study of
the wave functions of $^{124,126,128,130,132}$Xe, $^{124,126,128,130}$Te, $%
^{130,132}$Ba, $^{150}$Nd and $^{150}$Sm nuclei, calculating the yrast
spectra, reduced $B(E2$: $0^{+}\rightarrow 2^{+})$ transition probabilities,
static quadrupole moments $Q(2^{+})$ and $g$ -factors $g(2^{+})$ and
comparing them with the available experimental data. The half-lives $%
T_{1/2}^{2\nu }$ for the 2$\nu $ $\beta ^{-}\beta ^{-}$ decay of $^{128,130}$%
Te, $^{150}$Nd and 2$\nu $ e$^{+}$ DBD modes of $^{124,126}$Xe and $%
^{130,132}$Ba isotopes for the 0$^{+}\rightarrow $0$^{+}$ transition are
calculated. The role of deformation on NTMEs $M_{2\nu }$ is also studied. We
present some concluding remarks in Section 4.

\section{Theoretical framework}

The inverse half-life of the 2$\nu $ $\beta \beta $ decay modes for the $%
0^{+}\rightarrow 0^{+}$ transition is given by 
\begin{equation}
\left[ T_{1/2}^{2\nu }(k)\right] ^{-1}=G_{2\nu }(k)\left| M_{2\nu
}(k)\right| ^{2}
\end{equation}
where $k$ denotes the $\beta ^{-}\beta ^{-}$ and $e^{+}$DBD modes. The
integrated kinematical factor $\ G_{2\nu }(k)$\ can be calculated with good
accuracy \cite{doi92} and the NTME $M_{2\nu }(k)$ is defined as

\begin{equation}
M_{2\nu }(k)=\sum\limits_{N}\frac{\langle 0_{F}^{+}||\mathbf{\sigma }\tau
^{\pm }||1_{N}^{+}\rangle \langle 1_{N}^{+}||\mathbf{\sigma }\tau ^{\pm
}||0^{+}\rangle }{(E_{N}-E_{I}) + E_0} .  \label{m2n}
\end{equation}
where 
the total released energy $W_{0}(k) = 2 E_0$, for the different 2$\nu $ $%
\beta \beta $ decay modes, is given by 
\begin{eqnarray}
W_{0}(\beta ^{-}\beta ^{-}) &=&Q_{\beta ^{-}\beta ^{-}}+2m_{e} \\
W_{0}(\beta ^{+}\beta ^{+}) &=&Q_{\beta ^{+}\beta ^{+}}+2m_{e} \\
W_{0}(\beta ^{+}EC) &=&Q_{\beta ^{+}EC}+e_{b} \\
W_{0}(ECEC) &=&Q_{ECEC}-2m_{e}+e_{b1}+e_{b2}
\end{eqnarray}

The numerical evaluation of Eq. (\ref{m2n}) requires an explicit summation
over all the states of the intermediate odd $Z$-odd $N$ nuclei. However, it
is not possible to study the structure of intermediate odd-odd nuclei in the
present version of the PHFB model. Therefore, we carry out the summation
over intermediate states \cite{cha05,rai06} using the summation method given
by Civitarese \textit{et al.} \cite{civ93}. 
This technique is similar to the Operator Expansion Method \cite{wu91},
which has been proven to be in general inaccurate \cite{civ93,eng92}.
However, when the GT operator commutes with the effective two-body
interaction, the multiple commutators involved can be expressed as linear
functions of the single particle energies allowing for an exact sum over all
intermediate states \cite{cha05,cas94,hir95}. The Hamiltonian given by Eq.
(13) does not include spin dependent two-body terms, allowing for the use of
the summation method. While very convenient, and useful in the study of
electromagnetic transitions \cite{dix02}, the model is explicitly excluding
the spin-isospin interactions which are known to control the size of the 2$%
\nu $ $\beta ^{-}\beta ^{-}$ decay matrix elements in QRPA calculations \cite
{fae88,suh98,vog86}. Despite this strong limitation, PHFB calculations have
proved to be able to reproduce the observed 2$\nu $ $\beta ^{-}\beta ^{-}$
decay half-lives through the explicit introduction of deformation \cite
{cha05}. An extended version of the present model, able to describe odd-odd
nuclei, would be needed to include at the same time quadrupole-quadrupole
and spin-isospin two body terms in the Hamiltonian, and evaluate the $\beta
\beta $ decay matrix elements without employing the summation method. It
represents a technical challenge which exceeds the objectives of the present
work. 
Using the summation method, the NTME $M_{2\nu }(k)$ can be written as 
\begin{equation}
M_{2\nu }(k)=\frac{1}{E_{0}}\left\langle 0_{F}^{+}\left| \sum_{\mu
}(-1)^{\mu }\Gamma _{-\mu }F_{\mu }\right| 0_{I}^{+}\right\rangle
\label{m2nsu}
\end{equation}
where $\Gamma _{\mu }$ is given by 
\begin{equation}
\Gamma _{\mu }=\sigma _{\mu }\tau ^{\pm }
\end{equation}
and 
\begin{equation}
F_{\mu }=\sum_{\lambda =0}^{\infty }\frac{(-1)^{\lambda }}{E_{0}^{\lambda }}%
D_{\lambda }\Gamma _{\mu }
\end{equation}
with 
\begin{equation}
D_{\lambda }\Gamma _{\mu }=\left[ H,\left[ H,........,\left[ H,\Gamma _{\mu
}\right] .......\right] \right] ^{(\lambda \hbox{ times})}  \label{eqcom}
\end{equation}
Assuming that the GT operator commutes with the effective two-body
interaction, the Eq. (\ref{eqcom}) can be further simplified to 
\begin{equation}
M_{2\nu }(k)=\sum\limits_{\pi ,\nu }\frac{\langle 0_{F}^{+}||\left( \mathbf{%
\sigma .\sigma }\tau ^{\pm }\tau ^{\pm }\right) _{\pi \nu
}||0_{I}^{+}\rangle }{E_{0}+\Delta _{\pi \nu }\left( k\right) }
\label{m2nsum}
\end{equation}
where 
\begin{equation}
\Delta _{\pi \nu }\left( k\right) =\left\{ 
\begin{array}{llll}
\varepsilon (n_{\pi },l_{\pi },j_{\pi })-\varepsilon (n_{\nu },l_{\nu
},j_{\nu }) &  & \hbox{for} & \beta ^{-}\beta ^{-}\hbox{mode} \\ 
\varepsilon (n_{\nu },l_{\nu },j_{\nu })-\varepsilon (n_{\pi },l_{\pi
},j_{\pi }) &  & \hbox{for} & e^{+}\hbox{DBD modes}
\end{array}
\right.
\end{equation}
When a two-body isoscalar quadrupole-quadrupole interaction is employed, as
in the case of the pseudo SU(3) model [39-41], the energy denominator is a
well-defined quantity without any free parameter, because the GT operator
commutes with this two-body interaction. In the pseudo SU(3) scheme the sum
in the above equation reduces to a single term for 2$\nu $ $\beta ^{-}\beta
^{-}$ \cite{cas94,hir95} and 2$\nu $ e$^{+}$DBD modes \cite{cer99}.

In the present work, we use a Hamiltonian with \textit{PPQQ} type \cite
{bar68} of effective two-body interaction. The Hamiltonian is explicitly
written as 
\begin{equation}
H=H_{sp}+V(P)+\zeta _{qq}V(QQ)  \label{hmtn}
\end{equation}
where $H_{sp}$ denotes the single particle Hamiltonian. Further,$\ \zeta
_{qq}$ denotes the strength of \textit{QQ} part of the effective two-body
interaction. The purpose of introducing it is to study the role of
deformation by varying the strength parameter $\zeta _{qq}$. The final
results are obtained by setting the $\ \zeta _{qq}$ = 1. The pairing part of
the effective two-body interaction $V(P)$ is written as 
\begin{equation}
V{(}P{)}=-\left( \frac{G}{4}\right) \sum\limits_{\alpha \beta
}(-1)^{j_{\alpha }+j_{\beta }-m_{\alpha }-m_{\beta }}a_{\alpha }^{\dagger
}a_{\bar{\alpha}}^{\dagger }a_{\bar{\beta}}a_{\beta }
\end{equation}
where $\alpha $ denotes the quantum numbers (\textit{nljm}). The state $\bar{%
\alpha}$ is same as $\alpha $ but with the sign of \textit{m} reversed. The 
\textit{QQ} part of the effective interaction $V(QQ)$\ is expressed as 
\begin{equation}
V(QQ)=-\left( \frac{\chi }{2}\right) \sum\limits_{\alpha \beta \gamma \delta
}\sum\limits_{\mu }(-1)^{\mu }\langle \alpha |q_{\mu }^{2}|\gamma \rangle
\langle \beta |q_{-\mu }^{2}|\delta \rangle \ a_{\alpha }^{\dagger }a_{\beta
}^{\dagger }\ a_{\delta }\ a_{\gamma }
\end{equation}
where 
\begin{equation}
{q_{\mu }^{2}}=\left( \frac{16\pi }{5}\right) ^{1/2}r^{2}Y_{\mu }^{2}(\theta
,\phi )
\end{equation}
The model Hamiltonian given by Eq. (\ref{hmtn}) is not isospin symmetric.
Hence, the energy denominator is not as simple as in the case of pseudo
SU(3) model. However, the violation of isospin symmetry for the \textit{QQ}
part of our model Hamiltonian is negligible, as will be evident from the
parameters of the two-body interaction given later. Further, the violation
of isospin symmetry for the pairing part of the two-body interaction is
presumably small in the mass region under study. Under these assumptions,
the expression to calculate the NTME $M_{2\nu }$ of 2$\nu $ e$^{+}$DBD modes
for $0^{+}\rightarrow 0^{+}$ transition in the PHFB\ model can be obtained
by generalizing the above ideas, as follows.

In the PHFB model, states with good angular momentum $\mathbf{J}$ are
obtained from the axially symmetric HFB intrinsic state ${|\Phi _{0}\rangle }
$ with \textit{K}=0 using the standard projection technique \cite{oni66}
given by 
\begin{equation}
{|\Psi _{00}^{J}\rangle }=\left[ \frac{(2J+1)}{{8\pi ^{2}}}\right] \int
D_{00}^{J}(\Omega )R(\Omega )|\Phi _{0}\rangle d\Omega
\end{equation}
where $\ R(\Omega )$\ and $\ D_{00}^{J}(\Omega )$\ are the rotation operator
and the rotation matrix respectively. The axially symmetric HFB intrinsic
state ${|\Phi _{0}\rangle }$ can be written as 
\begin{equation}
{|\Phi _{0}\rangle }=\prod\limits_{im}(u_{im}+v_{im}b_{im}^{\dagger }b_{i%
\bar{m}}^{\dagger })|0\rangle
\end{equation}
where the creation operators $\ b_{im}^{\dagger }$\ and $\ b_{i\bar{m}%
}^{\dagger }$\ are defined as 
\begin{equation}
b{_{im}^{\dagger }}=\sum\limits_{\alpha }C_{i\alpha ,m}a_{\alpha m}^{\dagger
}\quad \hbox{and}\mathrm{\quad }b_{i\bar{m}}^{\dagger }=\sum\limits_{\alpha
}(-1)^{l+j-m}C_{i\alpha ,m}a_{\alpha ,-m}^{\dagger }
\end{equation}
The results of HFB calculations are summarized by the amplitudes $%
(u_{im},v_{im})$ and expansion coefficients $C_{ij,m}$, 
which are obtained through minimization of the expectation value of
Hamiltonian given by Eq. (13) before angular momentum projection. While a
Variation After Projection (VAP) has been shown to provide a more accurate
description of the energy spectra \cite{kho82}, they imply a numerical
effort which seems to be unnecessary given the ability of the present
version of the PHFB to reproduce the observed 2$\nu $ $\beta ^{-}\beta ^{-}$
decay matrix elements \cite{cha05}. 

Finally, one obtains the following expression for the NTME $M_{2\nu }(k)$ of
the 2$\nu $ $\beta \beta $ decay modes

\begin{eqnarray}
M_{2\nu }(k) &=&\sum\limits_{\pi ,\nu }\frac{\langle {\Psi _{00}^{J_{f}=0}}%
||\left( \mathbf{\sigma .\sigma }\tau ^{\pm }\tau ^{\pm } \right)_{\pi \nu}||%
{\Psi _{00}^{J_{i}=0}}\rangle }{E_{0}+\Delta _{\beta }\left( k\right) } 
\nonumber \\
&=&[n_{X}^{J_{i}=0}n_{Y}^{J_{f}=0}]^{-1/2}\int\limits_{0}^{\pi
}n_{(X,Y)}(\theta )\sum\limits_{\alpha \beta \gamma \delta }\frac{%
\left\langle \alpha \beta \left| \mathbf{\sigma }_{1}.\mathbf{\sigma }%
_{2}\tau ^{\pm }\tau ^{\pm }\right| \gamma \delta \right\rangle }{%
E_{0}+\Delta _{\beta }\left( k\right) }  \nonumber \\
&&\times \sum_{\varepsilon \eta }\frac{(f_{Y}^{(p)*})_{\varepsilon \beta }}{%
\left[ 1+F_{X}^{(p)}(\theta )f_{Y}^{(p)*}\right] _{\varepsilon \alpha }}%
\frac{(F_{X}^{(q)*})_{\eta \delta }}{\left[ 1+F_{X}^{(q)}(\theta
)f_{Y}^{(q)*}\right] _{\gamma \eta }}\sin \theta d\theta  \label{eqf}
\end{eqnarray}
where $p$ and $q$ in the last line refer to protons (neutrons) and neutrons
(protons) in the parent (X) and daughter (Y) nuclei, for the $\beta
^{-}\beta ^{-}$ (e$^{+}$DBD) mode, respectively. Further,

\begin{equation}
n^{J}=\int\limits_{0}^{\pi }\{\det [1+F^{(\pi )}(\theta )f^{(\pi )\dagger
}]\}^{1/2}\times \{\det [1+F^{(\nu )}(\theta )f^{(\nu )\dagger
}]\}^{1/2}d_{00}^{J}(\theta )\sin (\theta )d\theta
\end{equation}
and

\begin{equation}
n_{(X,Y),}(\theta )=\{\det [1+F_{X}^{(\pi )}(\theta )f_{Y}^{(\pi )\dagger
}]\}^{1/2}\times \{\det [1+F_{X}^{(\nu )}(\theta )f_{Y}^{(\nu )\dagger
}]\}^{1/2}
\end{equation}
The matrices $f_{Y}$\ \ and $F_{X}(\theta )\ $are given by

\begin{equation}
f_{Y}=\sum_{i}C_{ij_{\alpha },m_{\alpha }}C_{ij_{\beta },m_{\beta }}\left(
v_{im_{\alpha }}/u_{im_{\alpha }}\right) \delta _{m_{\alpha },-m_{\beta }}
\label{eq1}
\end{equation}
and 
\begin{equation}
\left[ F_{X}(\theta )\right] _{\alpha \beta }=\sum_{m_{\alpha }^{^{\prime
}}m_{\beta }^{^{\prime }}}d_{m_{\alpha },m_{\alpha }^{^{\prime
}}}^{j_{\alpha }}(\theta )d_{m_{\beta },m_{\beta }^{^{\prime }}}^{j_{\beta
}}(\theta )f_{j_{\alpha }m_{\alpha }^{^{\prime }},j_{\beta }m_{\beta
}^{^{\prime }}}  \label{eq2}
\end{equation}

The energy denominator is evaluated as follows. Within the present scheme
the difference in proton and neutron single particle energies with the same
quantum numbers explains the Isobaric Analog State energy, and is well
described by the difference in Coulomb energies between the initial and
intermediate nuclei. When appropriate, the spin-orbit energy splitting is
also added. Hence 
\begin{eqnarray}
\varepsilon (n_{\pi },l_{\pi },j_{\pi })-\varepsilon (n_{\nu },l_{\nu
},j_{\nu }) &=&\left\{ 
\begin{array}{llll}
\Delta _{C} &  & for & n_{\nu }=n_{\pi },l_{\nu }=l_{\pi },j_{\nu }=j_{\pi }
\\ 
\Delta _{C}+\Delta E_{s.o.splitting} &  & for & n_{\nu }=n_{\pi },l_{\nu
}=l_{\pi },j_{\nu }\neq j_{\pi }
\end{array}
\right.  \nonumber \\
\varepsilon (n_{\nu },l_{\nu },j_{\nu })-\varepsilon (n_{\pi },l_{\pi
},j_{\pi }) &=&\left\{ 
\begin{array}{llll}
\Delta _{C}-2E_{0} &  & for & n_{\nu }=n_{\pi },l_{\nu }=l_{\pi },j_{\nu
}=j_{\pi } \\ 
\Delta _{C}-2E_{0}+\Delta E_{s.o.splitting} &  & for & n_{\nu }=n_{\pi
},l_{\nu }=l_{\pi },j_{\nu }\neq j_{\pi }
\end{array}
\right.  \label{den}
\end{eqnarray}
where the Coulomb energy difference $\Delta _{C}$ is given by Bohr and
Mottelson \cite{boh98} 
\begin{equation}
\Delta _{C}=\frac{0.70}{A^{1/3}}\left[ \left( 2Z+1\right) -0.76\left\{
\left( Z+1\right) ^{4/3}-Z^{4/3}\right\} \right]
\end{equation}

It must be underlined that, in the present context, the use of the summation
method goes beyond the closure approximation, because each proton-neutron
excitation is weighted depending on its spin-flip or non-spin-flip
character. The explicit inclusion of the spin-orbit splitting in the energy
denominator, Eq. (\ref{den}), implies that it cannot be factorized out of
the sum in Eq. (\ref{m2n}). In this sense, employing the summation method in
conjunction with the PHFB formalism is richer than what was done in previous
application with the pseudo SU(3) model \cite{cas94,hir95}.

To calculate the NTME $M_{2\nu }$ for the 2$\nu $ $\beta \beta $ modes, the
matrices $f_{Y}$ and $[F_{X}(\theta )]_{\alpha \beta }$ are evaluated using
expressions given by Eqs. (\ref{eq1}) and (\ref{eq2}) respectively. The
required NTME $M_{2\nu }(k)$ are obtained using Eq. (\ref{eqf}) with 20
gaussian quadrature points in the range (0, $\pi $).

\section{Results\ and discussions}

The doubly even $^{100}$Sn nucleus is treated as an inert core with the
valence space spanned by 2\textit{s}$_{1/2,}$1\textit{d}$_{3/2}$, 1\textit{d}%
$_{5/2}$, 1\textit{f}$_{7/2},$ 0\textit{g}$_{7/2}$, 0\textit{h}$_{9/2}$ and 0%
\textit{h}$_{11/2}$ orbits for protons and neutrons. The set of single
particle energies (SPE's) used (in MeV) are $\varepsilon $(1\textit{d}$%
_{5/2} $)=0.0, $\varepsilon $(2\textit{s}$_{1/2}$)=1.4, $\varepsilon $(1%
\textit{d}$_{3/2}$)=2.0, $\varepsilon $(0\textit{g}$_{7/2}$)=4.0, $%
\varepsilon $(0\textit{h}$_{11/2}$)=6.5 (4.8 for $^{150}$Nd and $^{150}$Sm), 
$\varepsilon $(1\textit{f}$_{7/2}$)=12.0 (11.5 for $^{150}$Nd and $^{150}$%
Sm), $\varepsilon $(0\textit{h}$_{9/2}$)=12.5 (12.0 for $^{150}$Nd and $%
^{150}$Sm) for protons and neutrons. This set of SPE's, but for the $%
\varepsilon $(0\textit{h}$_{11/2}$), which is increased by 1.5 MeV, has been
employed in a number of successful variational model calculations for
nuclear properties in the mass region \textit{A}$\sim $130 \cite{ran97}. 
The constant difference between proton and neutron SPE's with the same
quantum numbers, due to the Coulomb interaction, is only relevant in the
evaluation of the energy denominators, given by Eq. (25). The spin-orbit
splitting $\Delta E_{s.o.splitting}$ is evaluated using the above mentioned
SPE's. 

The strengths of the pairing interaction $V(PP)$ is fixed through the
relation $G_{p}=G_{n}=35/A$ $MeV.$ The strengths of the like particle
components of the \textit{QQ} interaction are taken as: $\chi _{pp}=\chi
_{nn}=0.0105$ $MeV$ $b^{-4}$, where $b$ is the oscillator parameter. The
strength of proton-neutron (\textit{pn}) component of the \textit{QQ}
interaction $\chi _{pn}$ is varied so as to obtain the yrast spectra in
optimum agreement with the experimental data. The theoretical spectra is
taken to be the optimum if the excitation energy of the 2$^{+}$ state 
\textit{E}$_{2^{+}}$ is reproduced as closely as possible in comparison to
the experimental results. Thus, we fix $\chi _{pn}$ through the
experimentally available energy spectra for a given model space, SPE's, $%
G_{p}$, $G_{n}$ and $\chi _{pp}$. We present the values of $\chi _{pn}$ in
Table 1. These values of the strength of the \textit{QQ} interaction are
comparable to those suggested by Arima on the basis of an empirical analysis
of the effective two-body interactions \cite{ari81}. All these input
parameters are kept fixed throughout the calculation. 
Further, we have performed independent calculations for the parent and
daughter nuclei involved in the $\beta \beta $ decay, whose deformations are
in general different. 

\subsection{Yrast spectra and electromagnetic properties}

In Table 1, we display the theoretically calculated and experimentally
observed yrast spectra of $^{124,126,128,130,132}$Xe, $^{124,126,128,130}$Te
and $^{130,132}$Ba isotopes for $J^{\pi }=$2$^{+},$ 4$^{+}$and 6$^{+}$
states. All the experimentally observed $E_{2^{+}}$ \cite{sak84} energies
are reproduced with two significant digits. However, it is noticed that the
theoretical spectra is expanded in comparison to the experimental spectra.
This effect can be corrected by complementing the PHFB model with the VAP
prescription \cite{kho82}. However, our aim is to reproduce properties of
the low-lying 2$^{+}$ states. Hence, we have not attempted to invoke the VAP
prescription, which will unnecessarily complicate the calculations.

We present the theoretically calculated as well as the experimentally
observed reduced $\ B(E2$: $0^{+}\to 2^{+})$ transition probabilities \cite
{ram87}, static quadrupole moments $\ Q(2^{+})$ and the gyromagnetic factors 
$\ g(2^{+})$ \cite{rag89} in Table 2. In the case of $B(E2$: $0^{+}\to
2^{+}) $, only some representative experimental as well as adopted values 
\cite{ram01} are tabulated. The calculated reduced $B(E2$: $0^{+}\to 2^{+})$
transition probabilities are presented for three values of the effective
charges, $\ e_{eff}$ =0.40, 0.50 and 0.60, in columns 2 to 4, respectively.
The experimentally observed data are tabulated in column 5. It is observed
that the calculated $B(E2$: $0^{+}\to 2^{+})$ are in excellent agreement
with the observed results in all cases except for $^{126}$Xe and $^{150}$Sm
isotopes. The calculated and experimentally observed $B(E2$: $0^{+}\to
2^{+}) $ transition probability of $^{126}$Xe and $^{150}$Sm differ by 0.015
and 0.466 e$^{2}$b$^{2}$ respectively for the same $e_{eff}$. However, they
are quite close for $e_{eff}$=0.4.

The theoretically calculated $Q(2^{+})$\textit{\ }are displayed in columns 6
to 8 of Table 2, for the same effective charges as mentioned above. The
experimental $Q(2^{+})$ data \cite{rag89} are given in column 9 of the same
table. No experimental $Q(2^{+})$\textit{\ }results are available for $%
^{124,126,128,130}$Xe and $^{132}$Ba nuclei. The calculated and experimental 
$Q(2^{+})$ are in agreement for $^{150}$Nd and $^{150}$Sm nuclei. For rest
of the nuclei, the theoretical $Q(2^{+})$ results are quite off from the
observed values.

The $g$-factors $g(2^{+})$ are calculated with $g_{l}^{\pi }=$1.0, $%
g_{l}^{\nu }=$0.0, and $g_{s}^{\pi }=g_{s}^{\nu }=$0.60, and are presented
in the column 10 of Table 2. The available experimental \textit{g(}2$^{+}$%
\textit{)} data \cite{rag89} are given in column 11. The calculated and
experimentally observed$\ g(2^{+})$ are in good agreement for $^{124}$Te and 
$^{126,128}$Xe isotopes, whereas they are off by 0.052, 0.054, 0.065 and
0.013 nm only in case of $^{124}$Xe, $^{126}$Te, $^{130}$Ba and $^{130}$Xe
isotopes respectively. However, the calculated $g(2^{+})$ values of $%
^{128,130}$Te, $^{132}$Ba, $^{132}$Xe, $^{150}$Nd and $^{150}$Sm nuclei are
off from the experimental results.

In general, the overall agreement between the calculated and observed
electromagnetic properties is reasonably good. From this analysis we
conclude that the PHFB wave functions of $^{124,126,128,130,132}$Xe, $%
^{124,126,128,130}$Te, $^{130,132}$Ba, $^{150}$Nd and $^{150}$Sm nuclei,
generated by fixing $\chi _{pn}$ to reproduce the energy of the first $2^{+}$
state, provide a reasonable input for calculating $\beta \beta $ decay
half-lives. In the following, we present the results of the NTMEs $M_{2\nu }$%
, and the half-lives $T_{1/2}^{2\nu }$, of $^{128,130}$Te, $^{150}$Nd nuclei
for 2$\nu $ $\beta ^{-}\beta ^{-}$ decay, and $^{124,126}$Xe and $^{130,132}$%
Ba nuclei for 2$\nu $ e$^{+}$DBD modes for the$\ 0^{+}\to 0^{+}$ transition,
using the same PHFB wave functions.

\subsection{Decay rates}

In the case of 2$\nu $ $\beta ^{-}\beta ^{-}$ decay, the phase space factors
for the 0$^{+}\rightarrow $0$^{+}$ transition have been given by Doi \textit{%
et al.} \cite{doi85} and Boehm \textit{et al.} for $\ g_{A}$= 1.25 \cite
{boe92}. We take the phase phase space factors from Boehm \textit{et al. }%
\cite{boe92} for calculating half-lives of 2$\nu $ $\beta ^{-}\beta ^{-}$
decay of $^{128,130}$Te and $^{150}$Nd isotopes. We take the phase space
factors for the 2$\nu $ e$^{+}$DBD modes from Doi \textit{et al.} \cite
{doi92} when available. For the rest of 2$\nu $ e$^{+}$DBD emitters, we
calculate the phase space factors following the prescription of Doi \textit{%
et al.} \cite{doi92}, in the approximation $C_{1}=1.0,$ $C_{2}=0.0 $, $%
C_{3}=0.0$ and $R_{1,1}(\varepsilon )=R_{+1}(\varepsilon
)+R_{-1}(\varepsilon )=$1.0. The phase space integrals have been evaluated
for $\ g_{A}$= 1.25 and 1.261, for the 0$^{+}\rightarrow 0^{+}$ transition
of 2$\nu $ $\beta ^{-}\beta ^{-}$ decay and 2$\nu $ e$^{+}$DBD modes,
respectively \cite{doi92}. However, it is more justified to use the nuclear
matter value of $\ g_{A}$ around 1.0 in heavy nuclei. Hence, the theoretical 
$T_{1/2}^{2\nu }$ are calculated for both$\ g_{A}$=1.0 and 1.261(1.25) for 2$%
\nu $ e$^{+}$DBD (2$\nu $ $\beta ^{-}\beta ^{-}$ decay) modes.

\subsubsection{$2\nu $ $\beta ^{-}\beta ^{-}$ decay}

In Table 3, all the available experimental and theoretical results for 2$\nu 
$ $\beta ^{-}\beta ^{-}$ decay of $^{128,130}$Te and $^{150}$Nd isotopes are
presented. The 2$\nu $ $\beta ^{-}\beta ^{-}$ decay of $^{128,130}$Te
isotopes for the 0$^{+}\rightarrow 0^{+}$ transition has been observed
geochemically, and results from direct detection methods are also available
in the case of $^{130}$Te and $^{150}$Nd isotopes. In the same Table 3, we
also give the experimentally extracted values of NTMEs $M_{2\nu }$. In Table
4, we compile all the available experimental and theoretical results for 2$%
\nu $ e$^{+}$DBD modes of $^{124,126}$Xe, $^{130,132}$Ba isotopes along with
our calculated NTMEs \textit{M}$_{2\nu }$, and the corresponding half-lives $%
T_{1/2}^{2\nu }$ for the 0$^{+}\rightarrow 0^{+}$ transition.

In the case of $^{128}$Te isotope, we take the phase space factor $G_{2\nu
}(\beta ^{-}\beta ^{-})$ = 8.475$\times $10$^{-22}$ y$^{-1}$. In the PHFB
model, the calculated $M_{2\nu }$ is close to the NTMEs extracted from
experiments of Takaoka \textit{et al.} \cite{tak96}, Bernatovicz \textit{et
al.} \cite{ber92} and Barabash \cite{bar02} for $g_{A}=$1.0, while it is
close to those of Lin \textit{et al. }\cite{lin88} and Manuel \cite{man86}
for $g_{A}=$1.25. The NTMEs $M_{2\nu }$ calculated in SU(4)$_{\sigma \tau }$ 
\cite{rum98}, SSDH \cite{sem00} and MCM \cite{aun96} differ from $M_{2\nu }$
calculated in the present work by factor of 1.4 - 1.6 and are in agreement
with the extracted $M_{2\nu }$ due to Lin \textit{et al.} \cite{lin88} and
Manuel \cite{man86} for $g_{A}=$1.0. The NTME $M_{2\nu }$ calculated by
Civitarese \textit{et al.} \cite{civ98} in SSDH is close to the NTMEs
extracted from the experiments of Takaoka \textit{et al.} \cite{tak96},
Manuel \cite{man91} and Barabash \cite{bar02} for $g_{A}=$1.25. The
presently calculated NTME is smaller than the $M_{2\nu }$ calculated in QRPA 
\cite{eng88} and WCSM \cite{hax84} by factor of approximately 2.2 and 3.6
respectively while it is larger by a factor of approximately 5 than the $%
M_{2\nu }$ calculated by Stoica \cite{sto94} in SRPA(WS). The half-life $%
T_{1/2}^{2\nu }$ calculated in QRPA \cite{sta90} is close to the
experimental $T_{1/2}^{2\nu }$ due to Takaoka \textit{et al.} \cite{tak96},
Lin \textit{et al. }\cite{lin88} and Barabash \cite{bar02}. The calculated
half-lives $T_{1/2}^{2\nu }$ by Hirsch \textit{et al.} \cite{hir94a} in OEM
and Scholten \textit{et al. }\cite{sch85} in IBM are quite small. In the
present calculation, the predicted half-life $T_{1/2}^{2\nu }$ is (1.05-2.55)%
$\times 10^{24}$ y for $g_{A}$ = (1.25 - 1.0).

The phase space factor $G_{2\nu }(\beta ^{-}\beta ^{-})$ used to study the 2$%
\nu $ $\beta ^{-}\beta ^{-}$ decay of $^{130}$Te is 4.808$\times $10$^{-18}$
y$^{-1}$. The NTME $M_{2\nu }$ calculated in the PHFB model is in agreement
with the NTME extracted from the experiment of Milano+INFN experiment \cite
{arn03} for $g_{A}=$1.0. The NTME $M_{2\nu }$ calculated by Caurier \textit{%
et al.} \cite{cau99} in the SM and Aunola \textit{et al.} \cite{aun96} in
the MCM are close to the recommended value of Barabash \cite{bar02} for $%
g_{A}$ =1.0. The NTME $M_{2\nu }$ calculated by Stoica \textit{et al.} \cite
{sto94} in SRPA (WS) is in agreement with the experimental $M_{2\nu }$ of
Lin \textit{et al.} \cite{lin88}, Takaoka \textit{et al.} \cite{tak96},
Barabash \cite{bar02} and Milano+INFN experiment \cite{arn03} for $g_{A}=$%
1.25. The presently calculated NTME differs from the NTME calculated in SU(4)%
$_{\sigma \tau }$ \cite{rum98} and QRPA \cite{eng88} by a factor of 1.1 and
1.2 respectively. In the RQRPA, the calculated $M_{2\nu }$ by Toivanen 
\textit{et al.} \cite{toi97} is close to the experimental value of
Bernatovicz \textit{et al.} \cite{ber92} for $g_{A}=$1.25. The calculated
half-lives $T_{1/2}^{2\nu }$ by Hirsch \textit{et al.} \cite{hir94a} in OEM,
Scholten \textit{et al. }\cite{sch85} in IBM and Haxton \textit{et al. }\cite
{hax84} in WCSM are quite small while the half-life $T_{1/2}^{2\nu }$
calculated by \cite{rum95} in SU(4)$_{\sigma \tau }$ is close to the
experimental results of Lin \textit{et al.} \cite{lin88}, Takaoka \textit{et
al.} \cite{tak96} and Milano+INFN experiment \cite{arn03}. In the present
calculation, the predicted half-life $T_{1/2}^{2\nu }$ is (1.16-2.82)$\times
10^{20}$ y for $g_{A}$ = (1.25 - 1.0).

In the case of $^{150}$Nd, the phase space factor $G_{2\nu }(\beta ^{-}\beta
^{-})$ used is 1.189$\times $10$^{-16}$ y$^{-1}$. The experimental $M_{2\nu
} $ of UCI \cite{sil97} and NEMO 3 \cite{lal05} for $g_{A}$ =1.25 and
ITEP+INR \cite{art95} for $g_{A}$ =1.0 are close to the calculated value of $%
M_{2\nu } $ in present work. The $M_{2\nu }$ calculated by Hirsch \textit{et
al.} \cite{hir95} in pseudo-SU(3) is in close agreement to the experimental $%
M_{2\nu }$ of UCI \cite{sil97} and NEMO 3 \cite{lal05} for $g_{A}$ =1.0. The
calculated value of $M_{2\nu }$ by \cite{cas94} in pseudo-SU(3) and \cite
{rum98} in SU(4)$_{\sigma \tau }$ is larger by a factor of 1.67 and 1.95
respectively than the presently calculated value of $M_{2\nu }.$ The
half-life $T_{1/2}^{2\nu }$ calculated by Hirsch \textit{et al.} \cite
{hir94a} in the OEM is close to the experimental result of ITEP+INR
experiment while the $T_{1/2}^{2\nu }$ calculated by Staudt \textit{et al.} 
\cite{sta90} in QRPA favors the value of UCI experiment. The predicted
half-life $T_{1/2}^{2\nu }$ in our PHFB model is (7.89-19.27)$\times 10^{18}$
y for $g_{A}$ = (1.25 - 1.0).

\subsubsection{$2\nu $ e$^{+}$DBD modes}

In the case of $^{124}$Xe, the available experimental half-life limits on $%
T_{1/2}^{2\nu }$ of 2$\nu $ $\beta ^{+}\beta ^{+}$ and 2$\nu $ $ECEC$ modes,
for the $0^{+}\rightarrow 0^{+}$ transition, are $>8.0\times 10^{23}$ y \cite
{bol97} and $>1.1\times 10^{17}$ y \cite{gav98} respectively. Barabash \cite
{bar89} has investigated 2$\nu $ $\beta ^{+}\beta ^{+}$ and 2$\nu $ $\beta
^{+}EC$ modes and the half-life limits $T_{1/2}^{2\nu }$ are $>2.0\times
10^{14}$ y and $>4.8\times 10^{16}$ y respectively. The phase space factors
used in the present calculation are $G_{2\nu }(\beta ^{+}\beta ^{+})$ = 1.205%
$\times $10$^{-25}$ y$^{-1}$, $G_{2\nu }(\beta ^{+}EC)$ = 4.353$\times $10$%
^{-21}$ y$^{-1}$ and $G_{2\nu }(ECEC)$ = 5.101$\times $10$^{-20}$ y$^{-1}$.
The calculated NTMEs $M_{2\nu }$ in the PHFB and SU(4)$_{\sigma \tau }$ \cite
{rum98} models are close to each other for 2$\nu $ $\beta ^{+}EC $ and 2$\nu 
$ $ECEC$ modes. The presently calculated NTME $M_{2\nu }$ is larger than the
NTME due to Aunola \textit{et al.} \cite{aun96} in MCM by a factor of 5 and
7 approximately for 2$\nu $ $\beta ^{+}EC$ and 2$\nu $ $ECEC$ modes
respectively while the NTMEs $M_{2\nu }$ calculated in the QRPA model \cite
{hir94} are larger than that of PHFB\ model values by a factor of 1.5
approximately for all the three modes. The calculated half-life $%
T_{1/2}^{2\nu }$ by Staudt \textit{et al.} \cite{sta91} differ by two orders
of magnitude form the half-lives $T_{1/2}^{2\nu }$ calculated in the present
work for all modes. The theoretically calculated $T_{1/2}^{2\nu }$ are of
the order of 10$^{25-27}$ y, 10$^{22-24}$ y and 10$^{21-24}$ y for 2$\nu $ $%
\beta ^{+}\beta ^{+}$, 2$\nu $ $\beta ^{+}EC$ and 2$\nu $ $ECEC$ modes
respectively for $g_{A}=1.261-1.00$. The calculated half-lives $%
T_{1/2}^{2\nu }$ in the PHFB model are (3.015-7.624)$\times $10$^{27}$ y,
(8.347-21.11)$\times $10$^{22}$ y and (7.123-18.01)$\times $10$^{21}$ y for 2%
$\nu $ $\beta ^{+}\beta ^{+}$, 2$\nu $ $\beta ^{+}EC$ and 2$\nu $ $ECEC$
modes respectively for $g_{A}$=(1.26-1.0).

The e$^{+}$DBD modes of $^{126}$Xe isotope for the $0^{+}\rightarrow 0^{+}$
transition have not been studied so far either experimentally or
theoretically. In the present calculation, the predicted half-life $%
T_{1/2}^{2\nu }$ is (5.682-14.37)$\times $10$^{24}$ y for $g_{A}$ =
(1.261-1.0) using the phase space factor $G_{2\nu }(ECEC)$ = 7.428$\times $
10$^{-23}$ y$^{-1}$ for 2$\nu $ $ECEC$ mode.

In the case of $^{130}$Ba isotope, the $0^{+}\rightarrow 0^{+}$ transition
of e$^{+}$DBD modes has been investigated in geochemical experiments by
Meshik \cite{mes01} as well as Barabash \cite{bar96a} and combined limits
for all the e$^{+}$DBD modes are given. Meshik \cite{mes01} has given a
half-life limit $T_{1/2}^{2\nu }=(2.16\pm 0.52)\times 10^{21}$ y. The
half-life limit for $(2\nu +0\nu )$ decay mode given by Barabash \cite
{bar96a} is $>4.0\times 10^{21}$ y for all the e$^{+}$DBD modes. The phase
space factors used in the present calculation are $G_{2\nu }(\beta ^{+}\beta
^{+})$ $=1.211\times $10$^{-27}$ y$^{-1}$, $G_{2\nu }(\beta ^{+}EC)\ $= 1.387%
$\times $10$^{-21}$ y$^{-1}$ and $G_{2\nu }(ECEC)$ = 4.134$\times $10$^{-20}$
y$^{-1}$respectively. The theoretical NTME $M_{2\nu }$ calculated in the
PHFB and SU(4)$_{\sigma \tau }$ \cite{rum98} models differ by a factor of $%
\sim $1.4 for 2$\nu $ $\beta ^{+}EC$ and 2$\nu $ $ECEC$ modes. The NTME $%
M_{2\nu }$ calculated in the PHFB model is smaller than the NTMEs of Hirsch 
\textit{et al.} \cite{hir94} by a factor of 1.7 in the case of 2$\nu $ $%
\beta ^{+}\beta ^{+}$ mode while for 2$\nu $ $\beta ^{+}EC$ and 2$\nu $ $%
ECEC $ modes, the results differ by a factor of 2 approximately. On the
other hand, the calculated NTME in the present work is larger than the
calculated $M_{2\nu }$ due to Aunola \textit{et al.} \cite{aun96} in MCM by
a factor of 2.7 and 6 approximately for 2$\nu $ $\beta ^{+}EC$ and 2$\nu $ $%
ECEC$ modes respectively. The calculated half-life $T_{1/2}^{2\nu }$ by
Staudt \textit{et al.} \cite{sta91} in QRPA model for 2$\nu $ $\beta
^{+}\beta ^{+}$ mode is of the order of present calculation. The predicted $%
T_{1/2}^{2\nu }$ of 2$\nu $ $\beta ^{+}\beta ^{+}$, 2$\nu $ $\beta ^{+}EC$
and 2$\nu $ $ECEC$ modes in PHFB model are $(4.797-12.13)\times $10$^{29}$
y, $(4.188-10.59)\times $10$^{23}$ y and $(1.405-3.553)\times $10$^{22}$ y
respectively for $g_{A}=(1.26-1.0)$. The experimentally observed half-life
for the 0$^{+}\rightarrow $0$^{+}$ transition of $^{130}$Ba \cite{mes01} is
close to the theoretically predicted $T_{1/2}^{2\nu }$ of 2$\nu $ $ECEC$
mode.

For the $^{132}$Ba isotope, the deduced total half-life limit for of all the
e$^{+}$DBD modes $(2\nu +0\nu $ decay$)$ for the $0^{+}\rightarrow 0^{+}$
transition by Barabash \cite{bar96a} in geochemical experiment is $%
>3.0\times 10^{20}$ y. Meshik \cite{mes01} has also carried out the
geochemical experiment and has given a combined half-life limit $%
T_{1/2}^{2\nu }=(1.3\pm 0.9)\times 10^{21}$ y for all the e$^{+}$DBD modes.
No theoretical study has been done so far for the study of e$^{+}$DBD modes
of $^{132}$Ba. The phase space factor $G_{2\nu }(ECEC)$ used in the present
calculation is 6.706$\times $10$^{-23}$ y$^{-1}$. In the present
calculation, the predicted half-life $T_{1/2}^{2\nu }$ for the 2$\nu $ $ECEC$
mode is $(5.474-13.84)\times 10^{24}$ y for $g_{A}=(1.261-1.0)$.

\subsection{Deformation effect}

There are several possibilities to quantify the deformation parameter of the
nucleus. We take the quadrupole moment of the intrinsic state $\left\langle
Q_{0}^{2}\right\rangle $ (in arbitrary units) and the quadrupole deformation
parameter $\beta _{2}$ as a quantitative measure of the deformation. The
variation of the $\left\langle Q_{0}^{2}\right\rangle $, $\beta _{2}$ and $%
M_{2\nu }$ with respect to the change in the strength of the \textit{QQ}
interaction $\zeta _{qq}$ has been investigated to understand the role of
deformation on the NTME $M_{2\nu }$. In table 5, the quadrupole moment of
the intrinsic states $\left\langle Q_{0}^{2}\right\rangle $, deformation
parameter $\beta _{2}$ and the NTMEs $M_{2\nu }$ for different $\zeta _{qq}$
are presented. We calculate the deformation parameter with the same
effective charges as used in the calculation of reduced $B(E2$:$0^{+}\to
2^{+})$ transition probabilities.

It is noticed that the $\left\langle Q_{0}^{2}\right\rangle $ as well as $%
\beta _{2}$ increases in general as the $\zeta _{qq}$ is varied from 0 to
1.5 except a few anomalies. In all cases, it is found that the quadrupole
deformation parameter $\beta _{2}$ follows the same behavior as the
quadrupole moment of the intrinsic state $\left\langle
Q_{0}^{2}\right\rangle $ with respect to the change in $\zeta _{qq}$.
Further, there is an anticorrelation between the deformation parameter and
the NTME $M_{2\nu }$ in general but for a few exceptions.

We define a quantity $D_{2\nu }$ as the ratio of $M_{2\nu }$ at zero
deformation ($\zeta _{qq}=0$) and full deformation ($\zeta _{qq}=1$) to
quantify the effect of deformation on $M_{2\nu }.$ The ratio $D_{2\nu }$ is
given by 
\begin{equation}
D_{2\nu }=\frac{M_{2\nu }(\zeta _{qq}=0)}{M_{2\nu }(\zeta _{qq}=1)}.
\end{equation}
The ratios $D_{2\nu }$ are 3.63, 3.47, 4.26, 2.89, 4.66, 3.10 and 6.02 for $%
^{124}$Xe, $^{126}$Xe, $^{128}$Te, $^{130}$Te, $^{130}$Ba, $^{132}$Ba and $%
^{150}$Nd nuclei respectively. These values of $D_{2\nu }$ suggest that the $%
M_{2\nu }$ is quenched by a factor of 3 to 6 approximately in the mass
region $A\sim 120-150$ due to deformation effects. In contrast to other
models, the quenching of the NTMEs seems to be closely related with the
explicit inclusion of deformation effects. 

In $\beta \beta $ decay studies where deformation was included but no
angular momentum projection was performed, nuclear deformation was found to
be a mechanism of suppression of the 2$\nu $ $\beta \beta $ decay. In this
case, the $\beta \beta $ decay matrix elements are found to have maximum
values for about equal deformations of parent and daughter nuclei, and they
decrease rapidly when differences in deformations increase \cite{alv04}.
This deformation effect is different from the one reported in this work.
Further research is needed to relate these two approaches.

The quenching factors discussed above could be considered as a conservative
estimate of the uncertainties in the predicted NTMEs of $2\nu $ $\beta \beta 
$ modes, given the fact that many of the nuclei studied are in the
transitional region and do not display a well defined rotational spectrum.
Also the \textit{PPQQ} interaction employed in the present calculation is of
schematic nature. These uncertainties qualify both the present results,
those obtained with other models where deformation is not explicitly
considered, or where the rotational symmetry is not restored. The
uncertainties associated with the $0\nu $ $\beta \beta $ decay processes
would be expected to be far smaller than in the $2\nu $ $\beta \beta $ modes.

\section{Conclusions}

To summarize, we have built the PHFB wave functions of the parent and
daughter nuclei involved in $\beta \beta $ decay processes. Theoretical
results for the yrast spectra, reduced $B(E2$: $0^{+}\rightarrow 2^{+})$
transition probabilities, static quadrupole moments $Q(2^{+})$ and $g$%
-factors $g(2^{+})$ of $^{124,126,128,130,132}$Xe, $^{124,126,128,130}$Te, $%
^{130,132}$Ba, $^{150}$Nd and $^{150}$Sm nuclei were presented and compared
with the available experimental results. The same PHFB wave functions were
employed to calculate NTMEs $M_{2\nu }$ and half-lives \textit{T}$%
_{1/2}^{2\nu }$ of $^{124}$Xe (2$\nu $ $\beta ^{+}\beta ^{+}$, 2$\nu $ $%
\beta ^{+}EC$ and 2$\nu $ $ECEC$ modes), $^{126}$Xe (2$\nu $ $ECEC$ mode), $%
^{128}$Te (2$\nu $ $\beta ^{-}\beta ^{-}$ decay), $^{130}$Te (2$\nu $ $\beta
^{-}\beta ^{-}$ decay), $^{130}$Ba (2$\nu $ $\beta ^{+}\beta ^{+}$, 2$\nu $ $%
\beta ^{+}EC$ and 2$\nu $ $ECEC$ modes), $^{132}$Ba (2$\nu $ $ECEC$ mode)
and $^{150}$Nd (2$\nu $ $\beta ^{-}\beta ^{-} $ decay) isotopes. It was
shown that the proton-neutron part of the \textit{PPQQ} interaction, which
is responsible for triggering deformation in the intrinsic ground state,
plays an important role in the quenching of \textit{M}$_{2\nu }$, by a
factor of approximately 3 to 6, in the considered mass region $A\sim $%
120-150.

This work was partially supported by DAE-BRNS, India vide sanction No.
2003/37/14/BRNS/669, by Conacyt-M\'exico and DGAPA-UNAM.

\noindent \smallskip \textbf{Table 1:} Excitation energies (MeV) of $J^{\pi
}=2^{+},$ $4^{+}$ and $6^{+}$ yrast states of $^{124,126,128,130}$Te, $%
^{124,126,128,130,132}$Xe, and $^{130,132}$Ba, $^{150}$Nd and $^{150}$Sm
nuclei.

\begin{tabular}{l}
\end{tabular}

\noindent \noindent \noindent 
\begin{tabular}{llllllllllll}
\hline\hline
Nucleus & $\chi _{pn}$ & \multicolumn{2}{c}{Theory} & Exp.\cite{sak84} &  & 
& Nucleus & $\chi _{pn}$ & \multicolumn{2}{l}{Theory} & Exp.\cite{sak84} \\ 
\hline
&  &  &  &  &  &  &  &  &  &  &  \\ 
$^{124}$Xe & 0.03403 & $E_{2^{+}}$ & 0.3533 & 0.3540 &  &  & $^{124}$Te & 
0.0423 & $E_{2^{+}}$ & 0.6027 & 0.6028 \\ 
&  & $E_{4^{+}}$ & 1.1322 & 0.8787 &  &  &  &  & $E_{4^{+}}$ & 1.8695 & 
1.2488 \\ 
&  & $E_{6^{+}}$ & 2.2490 & 1.5482 &  &  &  &  & $E_{6^{+}}$ & 3.5851 & 
1.7470 \\ \hline
$^{126}$Xe & 0.0315 & $E_{2^{+}}$ & 0.3887 & 0.3886 &  &  & $^{126}$Te & 
0.03562 & $E_{2^{+}}$ & 0.6663 & 0.66634 \\ 
&  & $E_{4^{+}}$ & 1.2295 & 0.9419 &  &  &  &  & $E_{4^{+}}$ & 2.0026 & 
1.3613 \\ 
&  & $E_{6^{+}}$ & 2.4085 & 1.6349 &  &  &  &  & $E_{6^{+}}$ & 3.7708 & 
1.7755 \\ \hline
$^{128}$Te & 0.02715 & $E_{2^{+}}$ & 0.7436 & 0.7432 &  &  & $^{128}$Xe & 
0.0360 & $E_{2^{+}}$ & 0.4511 & 0.4429 \\ 
&  & $E_{4^{+}}$ & 2.0458 & 1.4971 &  &  &  &  & $E_{4^{+}}$ & 1.4263 & 
1.0329 \\ 
&  & $E_{6^{+}}$ & 3.7363 & 1.8111 &  &  &  &  & $E_{6^{+}}$ & 2.7976 & 
1.7370 \\ \hline
$^{130}$Te & 0.01801 & $E_{2^{+}}$ & 0.8393 & 0.8395 &  &  & $^{130}$Xe & 
0.02454 & $E_{2^{+}}$ & 0.5385 & 0.5361 \\ 
&  & $E_{4^{+}}$ & 1.7741 & 1.6325 &  &  &  &  & $E_{4^{+}}$ & 1.5496 & 
1.2046 \\ 
&  & $E_{6^{+}}$ & 3.0833 & 1.8145 &  &  &  &  & $E_{6^{+}}$ & 2.7831 & 
1.9444 \\ \hline
$^{130}$Ba & 0.0327 & $E_{2^{+}}$ & 0.3600 & 0.3574 &  &  & $^{130}$Xe & 
0.02454 & $E_{2^{+}}$ & 0.5385 & 0.5361 \\ 
&  & $E_{4^{+}}$ & 1.1605 & 0.9014 &  &  &  &  & $E_{4^{+}}$ & 1.5496 & 
1.2046 \\ 
&  & $E_{6^{+}}$ & 2.3208 & 1.5925 &  &  &  &  & $E_{6^{+}}$ & 2.7831 & 
1.9444 \\ \hline
$^{132}$Ba & 0.01302 & $E_{2^{+}}$ & 0.4645 & 0.4646 &  &  & $^{132}$Xe & 
0.01536 & $E_{2^{+}}$ & 0.6675 & 0.6677 \\ 
&  & $E_{4^{+}}$ & 1.4519 & 1.1277 &  &  &  &  & $E_{4^{+}}$ & 1.6278 & 
1.4403 \\ 
&  & $E_{6^{+}}$ & 2.7982 & 1.9328 &  &  &  &  & $E_{6^{+}}$ & 2.7263 & 
2.1118 \\ \hline
$^{150}$Nd & 0.02160 & $E_{2^{+}}$ & 0.1307 & 0.13012 &  &  & $^{150}$Sm & 
0.01745 & $E_{2^{+}}$ & 0.3328 & 0.33395 \\ 
&  & $E_{4^{+}}$ & 0.4320 & 0.3815 &  &  &  &  & $E_{4^{+}}$ & 1.0156 & 
0.77335 \\ 
&  & $E_{6^{+}}$ & 0.8960 &  &  &  &  &  & $E_{6^{+}}$ & 1.9185 & 1.27885 \\ 
\hline\hline
\end{tabular}

\begin{tabular}{l}
\end{tabular}

\pagebreak

\noindent \nolinebreak \textbf{Table 2:} Comparison of theoretically
calculated and experimentally observed reduced transition probabilities
\noindent \textit{B}(\textit{E2: $0^{+}\rightarrow 2^{+}$}), static
quadrupole moments \ \textit{Q}(\textit{2$^{+}$}) and $g$ factors \textit{g}(%
\textit{\ 2$^{+}$}) of $^{124,126,128,130}$Te, $^{124,126,128,130,132}$Xe,
and $^{130,132}$Ba nuclei. Here \textit{B}(\textit{E2}) and \textit{Q}(%
\textit{2$^{+}$}) are calculated for effective charge e$_{p}=$1+e$_{eff}$
and e$_{n}=$e$_{eff}$. $^{*}$ denotes the average \textit{B}(\textit{E2})
values.

\noindent 
\begin{tabular}{l}
\end{tabular}

\noindent 
\begin{tabular}{llllclllccl}
\hline\hline
Nucleus & \multicolumn{4}{c}{$B(E2:0^{+}\rightarrow 2^{+})$ (e$^{2}$b$^{2}$)}
& \multicolumn{4}{c}{\ \ \ $Q(2^{+})$ (eb)} & \multicolumn{2}{c}{\ $g(2^{+})$
(nm)} \\ 
& \multicolumn{3}{c}{\ Theory} & \ Exp.\cite{ram87} & \multicolumn{3}{c}{\ \
Theory} & \ Exp.\cite{rag89} & Theory & Exp.\cite{rag89} \\ 
&  & e$_{eff}$ &  & \multicolumn{1}{l}{} &  & e$_{eff}$ &  &  &  &  \\ 
\cline{2-4}\cline{6-8}
& 0.40 & 0.50 & 0.60 & \multicolumn{1}{l}{} & 0.40 & 0.50 & 0.60 & 
\multicolumn{1}{l}{} & \multicolumn{1}{l}{} &  \\ \hline
&  &  &  &  &  &  &  &  &  &  \\ 
$^{124}$Xe & 0.722 & \textbf{0.941} & 1.188 & \multicolumn{1}{l}{0.96$\pm
0.06^{*}$} & -0.770 & \textbf{-0.879} & -0.988 & \multicolumn{1}{l}{-} & 
\multicolumn{1}{l}{0.302} & 0.23$\pm 0.02$ \\ 
&  &  &  & \multicolumn{1}{l}{0.90$\pm 0.07$} &  &  &  & \multicolumn{1}{l}{}
& \multicolumn{1}{l}{} &  \\ 
&  &  &  & \multicolumn{1}{l}{1.49$\pm 0.09$} &  &  &  & \multicolumn{1}{l}{}
& \multicolumn{1}{l}{} &  \\ \hline
$^{124}$Te & 0.401 & \textbf{0.530} & 0.677 & \multicolumn{1}{l}{0.568$\pm
0.006^{*}$} & -0.574 & \textbf{-0.659} & -0.745 & \multicolumn{1}{l}{-0.45$%
\pm 0.05$} & \multicolumn{1}{l}{0.348} & 0.33$\pm 0.03$ \\ 
&  &  &  & \multicolumn{1}{l}{0.61$\pm 0.20$} &  &  &  & \multicolumn{1}{l}{}
& \multicolumn{1}{l}{} & 0.31$\pm 0.04$ \\ 
&  &  &  & \multicolumn{1}{l}{0.538$\pm 0.028$} &  &  &  & 
\multicolumn{1}{l}{} & \multicolumn{1}{l}{} & 0.28$\pm 0.03$ \\ \hline
$^{126}$Xe & 0.670 & \textbf{0.865} & 1.084 & \multicolumn{1}{l}{0.770$\pm
0.025^{*}$} & -0.742 & \textbf{-0.843} & -0.943 & \multicolumn{1}{l}{-} & 
\multicolumn{1}{l}{0.373} & 0.37$\pm 0.07$ \\ 
&  &  &  & \multicolumn{1}{l}{0.79$\pm 0.06$} &  &  &  & \multicolumn{1}{l}{}
& \multicolumn{1}{l}{} & 0.27$\pm 0.04$ \\ 
&  &  &  & \multicolumn{1}{l}{0.760$\pm 0.026$} &  &  &  & 
\multicolumn{1}{l}{} & \multicolumn{1}{l}{} &  \\ \hline
$^{126}$Te & 0.353 & \textbf{0.460} & 0.581 & \multicolumn{1}{l}{0.475$\pm
0.010^{*}$} & -0.539 & \textbf{-0.614} & -0.691 & \multicolumn{1}{l}{-0.20$%
\pm 0.09$} & \multicolumn{1}{l}{0.424} & 0.34$\pm 0.03$ \\ 
&  &  &  & \multicolumn{1}{l}{0.487$\pm 0.035$} &  &  &  & 
\multicolumn{1}{l}{} & \multicolumn{1}{l}{} & 0.31$\pm 0.04$ \\ 
&  &  &  & \multicolumn{1}{l}{0.532$\pm 0.037$} &  &  &  & 
\multicolumn{1}{l}{} & \multicolumn{1}{l}{} &  \\ \hline
$^{128}$Te & 0.298 & \textbf{0.381} & 0.474 & \multicolumn{1}{l}{0.383$\pm $%
0.006$^{*}$} & -0.496 & \textbf{-0.561} & -0.626 & \multicolumn{1}{l}{-0.06$%
\pm $0.05} & \multicolumn{1}{l}{0.514} & 0.35$\pm $0.04 \\ 
&  &  &  & \multicolumn{1}{l}{0.380$\pm $0.009} &  &  &  & 
\multicolumn{1}{l}{-0.14$\pm $0.12} & \multicolumn{1}{l}{} & 0.31$\pm $0.04
\\ 
&  &  &  & \multicolumn{1}{l}{0.387$\pm $0.011} &  &  &  & 
\multicolumn{1}{l}{} & \multicolumn{1}{l}{} &  \\ \hline
$^{128}$Xe & 0.637 & \textbf{0.819} & 1.024 & \multicolumn{1}{l}{0.750$\pm $%
0.040$^{*}$} & -0.724 & \textbf{-0.820} & -0.917 & \multicolumn{1}{l}{} & 
\multicolumn{1}{l}{0.400} & 0.41$\pm $0.07 \\ 
&  &  &  & \multicolumn{1}{l}{0.790$\pm $0.040} &  &  &  & 
\multicolumn{1}{l}{} & \multicolumn{1}{l}{} & 0.31$\pm $0.03 \\ 
&  &  &  & \multicolumn{1}{l}{0.890$\pm $0.230} &  &  &  & 
\multicolumn{1}{l}{} & \multicolumn{1}{l}{} &  \\ \hline
$^{130}$Te & 0.231 & \textbf{0.289} & 0.354 & \multicolumn{1}{l}{0.295$\pm $%
0.007$^{*}$} & -0.438 & \textbf{-0.490} & -0.542 & \multicolumn{1}{l}{-0.15$%
\pm $0.10} & \multicolumn{1}{l}{0.679} & 0.33$\pm $0.08 \\ 
&  &  &  & \multicolumn{1}{l}{0.290$\pm $0.011} &  &  &  & 
\multicolumn{1}{l}{} & \multicolumn{1}{l}{} & 0.29$\pm $0.06 \\ 
&  &  &  & \multicolumn{1}{l}{0.260$\pm $0.050} &  &  &  & 
\multicolumn{1}{l}{} & \multicolumn{1}{l}{} &  \\ \hline
$^{130}$Xe & 0.493 & \textbf{0.624} & 0.769 & \multicolumn{1}{l}{0.65$\pm $%
0.05$^{*}$} & -0.637 & \textbf{-0.716} & -0.795 & \multicolumn{1}{l}{} & 
\multicolumn{1}{l}{0.463} & 0.38$\pm $0.07 \\ 
&  &  &  & \multicolumn{1}{l}{0.631$\pm $0.048} &  &  &  & 
\multicolumn{1}{l}{} & \multicolumn{1}{l}{} & 0.31$\pm $0.04 \\ 
&  &  &  & \multicolumn{1}{l}{0.640$\pm $0.160} &  &  &  & 
\multicolumn{1}{l}{} & \multicolumn{1}{l}{} &  \\ \hline
$^{130}$Ba & 1.048 & \textbf{1.331} & 1.649 & \multicolumn{1}{l}{1.163$\pm
0.016^{*}$} & -0.928 & \textbf{-1.046} & -1.164 & \multicolumn{1}{l}{-0.86$%
\pm 0.08$} & \multicolumn{1}{l}{0.445} & 0.35$\pm 0.03$ \\ 
&  &  &  & \multicolumn{1}{l}{1.36$\pm 0.14$} &  &  &  & \multicolumn{1}{l}{
-0.33$\pm 0.24$} & \multicolumn{1}{l}{} &  \\ 
&  &  &  & \multicolumn{1}{l}{1.21$\pm 0.38$} &  &  &  & \multicolumn{1}{l}{}
& \multicolumn{1}{l}{} &  \\ \hline
$^{132}$Ba & 0.624 & \textbf{0.767} & 0.926 & \multicolumn{1}{l}{0.86$\pm
0.06^{*}$} & -0.718 & \textbf{-0.796} & -0.874 & \multicolumn{1}{l}{-} & 
\multicolumn{1}{l}{0.649} & 0.34$\pm 0.03$ \\ 
&  &  &  & \multicolumn{1}{l}{0.73$\pm 0.18$} &  &  &  & \multicolumn{1}{l}{}
& \multicolumn{1}{l}{} &  \\ 
&  &  &  & \multicolumn{1}{l}{0.86$\pm 0.06$} &  &  &  & \multicolumn{1}{l}{}
& \multicolumn{1}{l}{} &  \\ \hline
$^{132}$Xe & 0.329 & \textbf{0.409} & 0.498 & \multicolumn{1}{l}{0.460$\pm
0.030^{*}$} & -0.519 & \textbf{-0.579} & -0.639 & \multicolumn{1}{l}{0.010$%
\pm 0.005$} & \multicolumn{1}{l}{0.575} & 0.39$\pm 0.05$ \\ 
&  &  &  & \multicolumn{1}{l}{0.42$\pm 0.11$} &  &  &  & \multicolumn{1}{l}{}
& \multicolumn{1}{l}{} & 0.37$\pm 0.05$ \\ 
&  &  &  & \multicolumn{1}{l}{0.44$\pm 0.03$} &  &  &  & \multicolumn{1}{l}{}
& \multicolumn{1}{l}{} &  \\ \hline
$^{150}$Nd & 2.132 & \textbf{2.580} & 3.070 & \multicolumn{1}{l}{2.760$\pm
0.040^{*}$} & -1.322 & \textbf{-1.455} & -1.587 & \multicolumn{1}{l}{-2.00$%
\pm 0.51$} & \multicolumn{1}{l}{0.636} & 0.422$\pm 0.039$ \\ 
&  &  &  & \multicolumn{1}{l}{2.640$\pm 0.080$} &  &  &  & 
\multicolumn{1}{l}{} & \multicolumn{1}{l}{} & 0.322$\pm 0.009$ \\ 
&  &  &  & \multicolumn{1}{l}{2.670$\pm 0.100$} &  &  &  & 
\multicolumn{1}{l}{} & \multicolumn{1}{l}{} &  \\ \hline
$^{150}$Sm & 1.707 & \textbf{2.056} & 2.437 & \multicolumn{1}{l}{1.350$\pm
0.030^{*}$} & -1.182 & \textbf{-1.297} & -1.412 & \multicolumn{1}{l}{-1.32$%
\pm 0.19$} & \multicolumn{1}{l}{0.592} & 0.385$\pm 0.027$ \\ 
&  &  &  & \multicolumn{1}{l}{1.470$\pm 0.090$} &  &  &  & 
\multicolumn{1}{l}{-1.25$\pm 0.20$} & \multicolumn{1}{l}{} & 0.411$\pm 0.032$
\\ 
&  &  &  & \multicolumn{1}{l}{1.440$\pm 0.150$} &  &  &  & 
\multicolumn{1}{l}{} & \multicolumn{1}{l}{} &  \\ \hline\hline
\end{tabular}

\noindent $^{*}$data taken from Reference \cite{ram01}

\begin{tabular}{l}
\end{tabular}

\pagebreak

\noindent \noindent

\noindent \textbf{Table 3:} Experimental half-lives \textit{T}$_{1/2}^{2\nu
} $ and corresponding matrix element \textit{M}$_{2\nu }$ of 2$\nu $ $\beta
^{-}\beta ^{-}$ decay for the $0^{+}\rightarrow 0^{+}$ transition of $%
^{128,130}$Te and $^{150}$Nd nuclei along with the theoretically calculated 
\textit{M}$_{2\nu }$ in different models. The numbers corresponding to (a)
and (b) are calculated for \textit{g}$_{A}=1.25$ and 1.0 respectively.

$
\begin{array}{l}
\end{array}
$

\noindent \noindent 
\begin{tabular}{lllllllllll}
\hline\hline
Nuclei & \multicolumn{5}{c}{Experiment} & \multicolumn{5}{c}{Theory} \\ 
& Ref. & Project & $T_{1/2}^{2\nu }$ &  & $\left| M_{2\nu }\right| $ & Ref.
& Model & $\left| M_{2\nu }\right| $ &  & $T_{1/2}^{2\nu }$ \\ \hline
&  &  &  &  &  &  &  &  &  &  \\ 
$^{128}$Te & \cite{tak96} & gch. & (2.2$\pm $0.3) & (a) & 0.023$%
_{-0.0014}^{+.00018}$ & * & PHFB & 0.033 & (a) & 1.05 \\ 
(10$^{24}$ y) &  &  &  & (b) & 0.036$_{-0.0022}^{+0.0027}$ &  &  &  & (b) & 
2.55 \\ 
& \cite{man91} & gch. & 2.0 & (a) & 0.024 & \cite{sem00} & SSDH & 0.048 & (a)
& 0.51 \\ 
&  &  &  & (b) & 0.038 &  &  &  & (b) & 1.29 \\ 
& \cite{ber92} & gch. & 7.7$\pm $0.4 & (a) & 0.012$_{-0.0003}^{+0.0003}$ & 
\cite{cau99} & SM & - &  & 0.5 \\ 
&  &  &  & (b) & 0.019$_{-0.0005}^{+0.0005}$ & \cite{rum98} & SU(4)$_{\sigma
\tau }$ & 0.053 & (a) & 0.42 \\ 
& \cite{lin88} & gch. & 1.8$\pm $0.7 & (a) & 0.026$_{-0.0039}^{+0.0071}$ & 
&  &  & (b) & 1.06 \\ 
&  &  &  & (b) & 0.040$_{-0.0060}^{+0.0111}$ & \cite{civ98} & SSDH & 0.013 & 
(a) & 6.98 \\ 
& \cite{kir86} & gch. & $>$5.0 & (a) & $<$0.015 &  &  &  & (b) & 17.65 \\ 
&  &  &  & (b) & $<$0.024 & \cite{aun96} & MCM & 0.046 & (a) & 0.56 \\ 
& \cite{man86} & gch. & 1.4$\pm $0.4 & (a) & 0.029$_{-0.0034}^{+0.0053}$ & 
&  &  & (b) & 1.41 \\ 
&  &  &  & (b) & 0.045$_{-0.0053}^{+0.0083}$ & \cite{sto94} & SRPA(WS) & 
0.006 & (a) & 32.78 \\ 
& \cite{ell02} & Average & 7.2$\pm 0.3$ & (a) & 0.013$_{-0.0003}^{+0.0003}$
&  &  &  & (b) & 82.87 \\ 
&  & Value &  & (b) & 0.020$_{-0.0004}^{+0.0004}$ & \cite{hir94a} & OEM & -
&  & 0.21 \\ 
& \cite{bar02} & Recommended & 2.5$\pm $0.4 & (a) & 0.022$%
_{-0.0015}^{+0.0020}$ & \cite{sta90} & QRPA & - &  & 2.63 \\ 
&  & Value &  & (b) & 0.034$_{-0.0024}^{+0.0031}$ & \cite{eng88} & QRPA & 
0.074 & (a) & 0.22 \\ 
&  &  &  &  &  &  &  &  & (b) & 0.54 \\ 
&  &  &  &  &  & \cite{sch85} & IBM & - &  & 0.09 \\ 
&  &  &  &  &  & \cite{hax84} & WCSM & 0.120 & (a) & 0.08 \\ 
&  &  &  &  &  &  &  &  & (b) & 0.21 \\ 
&  &  &  &  &  &  &  &  &  &  \\ 
$^{130}$Te & \cite{arn03} & Milano+INFN & 6.1$\pm $1.4$_{-3.5}^{+2.9}$ & (a)
& 0.018$_{-0.0043}^{+0.0232}$ & * & PHFB & 0.042 & (a) & 1.16 \\ 
(10$^{20}$ yr) &  &  &  & (b) & 0.029$_{-0.0068}^{+0.0362}$ &  &  &  & (b) & 
2.82 \\ 
& \cite{all00} & Milano & $>$3.0 & (a) & $<$0.026 & \cite{cau99} & SM & 0.030
& (a) & 2.3 \\ 
&  &  &  & (b) & $<$0.041 &  &  &  & (b) & 5.84 \\ 
& \cite{tak96} & gch. & 7.9$\pm $1.0 & (a) & 0.016$_{-0.0009}^{+0.0011}$ & 
\cite{rum98} & SU(4)$_{\sigma \tau }$ & 0.0468 & (a) & 0.95 \\ 
&  &  &  & (b) & 0.025$_{-0.0015}^{+0.0018}$ &  &  &  & (b) & 2.32 \\ 
& \cite{ber92} & gch. & 27.0$\pm $1.0 & (a) & 0.009$_{-0.0002}^{+0.0002}$ & 
\cite{toi97} & RQRPA(AWS) & 0.009 & (a) & 25.68 \\ 
&  &  &  & (b) & 0.014$_{-0.0002}^{+0.0003}$ &  &  &  & (b) & 62.70 \\ 
& \cite{man91} & gch. & 8.0 & (a) & 0.016 & \cite{toi97} & RQRPA(WS ) & 0.009
& (a) & 25.68 \\ 
&  &  &  & (b) & 0.025 &  &  &  & (b) & 62.70 \\ 
& \cite{lin88} & gch. & 7.5$\pm $0.3 & (a) & 0.017$_{-0.0003}^{+0.0003}$ & 
\cite{aun96} & MCM & 0.028 & (a) & 2.65 \\ 
&  &  &  & (b) & 0.026$_{-0.0005}^{+0.0005}$ &  &  &  & (b) & 6.48 \\ 
& \cite{bel87} &  & $>$8.0 & (a) & $<$ 0.016 & \cite{rum95} & SU(4)$_{\sigma
\tau }$ &  &  & 7.0 \\ 
&  &  &  & (b) & $<$0.025 & \cite{sto94} & SRPA(WS) & 0.016 & (a) & 8.12 \\ 
& \cite{zde80} & INR Kiev & $>$0.0001 & (a) & $<$4.563 &  &  &  & (b) & 19.84
\\ 
&  &  &  & (b) & $<$7.130 & \cite{hir94a} & OEM &  &  & 0.79 \\ 
& \cite{ell02} & Average & 27$\pm 1.0$ & (a) & 0.009$_{-0.0002}^{+0.0002}$ & 
\cite{sta90} & QRPA &  &  & 18.4 \\ 
&  & Value &  & (b) & 0.014$_{-0.0002}^{+0.0003}$ & \cite{eng88} & QRPA & 
0.049 & (a) & 0.87 \\ 
& \cite{bar02} & Recommended & 9.0$\pm $1.5 & (a) & 0.015$%
_{-0.0011}^{+0.0015}$ &  &  &  & (b) & 2.12 \\ 
&  & Value &  & (b) & 0.024$_{-0.0018}^{+0.0023}$ & \cite{sch85} & IBM &  & 
& 0.17 \\ 
&  &  &  &  &  & \cite{hax84} & WCSM & 0.114 & (a) & 0.16 \\ 
&  &  &  &  &  &  &  &  & (b) & 0.40 \\ \hline
\end{tabular}
\pagebreak

\begin{tabular}{lllllllllll}
\hline
\multicolumn{11}{l}{Table 3 continued} \\ \hline
$^{150}$Nd & \cite{lal05} & NEMO 3 & 9.7$\pm $0.7$\pm $1.0 & (a) & 0.029$%
_{-0.0023}^{+0.0030}$ & * & PHFB & 0.033 & (a) & 7.89 \\ 
(10$^{18}$ y) &  &  &  & (b) & 0.046$_{-0.0036}^{+0.0047}$ &  &  &  & (b) & 
19.27 \\ 
& \cite{sil97} & UCI & 6.75$_{-0.42}^{+0.37}\pm 0.68$ & (a) & 0.035$%
_{-0.0025}^{+0.0033}$ & \cite{rum98} & SU(4)$_{\sigma \tau }$ & 0.0642 & (a)
& 2.04 \\ 
&  &  &  & (b) & 0.055$_{-0.0038}^{+.00051}$ &  &  &  & (b) & 4.98 \\ 
& \cite{art95} & ITEP +INR & 18.8$_{-3.9}^{+6.6}\pm 1.9$ & (a) & 0.021$%
_{-0.0036}^{+0.0043}$ & \cite{cas94} & pSU(3) & 0.055 & (a) & 2.78 \\ 
&  &  &  & (b) & 0.033$_{-0.0056}^{+0.0067}$ &  &  &  & (b) & 6.79 \\ 
& \cite{vas90} &  & $>$11.0 & (a) & $<$0.028 & \cite{hir94a} & OEM &  &  & 
16.6 \\ 
&  &  &  & (b) & $<$0.043 & \cite{sta90} & QRPA &  &  & 7.37 \\ 
& \cite{art93} & ITEP +INR & 17$_{-5.0}^{+10}\pm $3.5 & (a) & 0.022$%
_{-0.0056}^{+0.0092}$ &  &  &  &  &  \\ 
&  &  &  & (b) & 0.035$_{-0.0088}^{+0.0144}$ &  &  &  &  &  \\ 
& \cite{ell93} & UCI & 9.0 & (a) & 0.031 &  &  &  &  &  \\ 
&  &  &  & (b) & 0.048 &  &  &  &  &  \\ 
& \cite{kli86} & INR & $>$18 & (a) & $<$0.022 &  &  &  &  &  \\ 
&  &  &  & (b) & $<$0.034 &  &  &  &  &  \\ 
& \cite{ell02} & Average & 7.0$_{-0.3}^{+11.8}$ & (a) & 0.035$%
_{-0.0135}^{+0.0008}$ &  &  &  &  &  \\ 
&  & Value &  & (b) & 0.054$_{-0.0211}^{+0.0012}$ &  &  &  &  &  \\ 
& \cite{bar02} & Average & 7.0$\pm $1.7 & (a) & 0.035$_{-0.0036}^{+0.0052}$
&  &  &  &  &  \\ 
&  & Value &  & (b) & 0.054$_{-0.0056}^{+0.0087}$ &  &  &  &  &  \\ 
&  &  &  &  &  &  &  &  &  &  \\ \hline\hline
\end{tabular}

gch. denotes the geochemical experiment, * denotes the present work

\noindent 
\begin{tabular}{l}
\end{tabular}

\pagebreak

\noindent \textbf{Table 4:} Experimental limits on half-lives $T_{1/2}^{2\nu
}(0^{+}\rightarrow 0^{+})$, theoretically calculated $M_{2\nu }$ and
corresponding $T_{1/2}^{2\nu }(0^{+}\rightarrow 0^{+})$ for 2$\nu $ e$^{+}$%
DBD modes of $^{124,126}$Xe and $^{130,132}$Ba nuclei. The numbers
corresponding to (a) and (b) are calculated for $g_{A}$=1.261 and 1.0
respectively.

\begin{tabular}{l}
\end{tabular}

\noindent \noindent 
\begin{tabular}{llllllllll}
\hline\hline
Nuclei & Decay & \multicolumn{2}{c}{Experiment} &  & \multicolumn{5}{c}{
Theory} \\ 
& Mode & Ref & $T_{1/2}^{2\nu }$ (y) &  & Ref. & Model & $\left| M_{2\nu
}\right| $ &  & $T_{1/2}^{2\nu }$ (y) \\ \hline
$^{124}$Xe & $\beta ^{+}\beta ^{+}$ & \cite{bol97} & $>$8.0$\times 10^{23}$
&  & * & PHFB & 0.0525 & (a) & 3.015$\times 10^{27}$ \\ 
&  & \cite{bar89} & $>$2.0$\times 10^{14}$ &  &  &  &  & (b) & 7.624$\times
10^{27}$ \\ 
&  &  &  &  & \cite{hir94} & QRPA & 0.0770 & (a) & 1.400$\times 10^{27}$ \\ 
&  &  &  &  &  &  &  & (b) & 3.539$\times 10^{27}$ \\ 
&  &  &  &  & \cite{sta91} & QRPA &  &  & 8.170$\times 10^{25}$ \\ 
\cline{2-10}
& $\beta ^{+}EC$ & \cite{bar89} & $>$4.8$\times 10^{16}$ &  & * & PHFB & 
0.0525 & (a) & 8.347$\times 10^{22}$ \\ 
&  &  &  &  &  &  &  & (b) & 2.111$\times 10^{23}$ \\ 
&  &  &  &  & \cite{rum98} & SU(4)$_{\sigma \tau }$ & 0.0528 & (a) & 8.240$%
\times 10^{22}$ \\ 
&  &  &  &  &  &  &  & (b) & 2.084$\times 10^{23}$ \\ 
&  &  &  &  & \cite{aun96} & MCM & 0.0100 & (a) & 2.297$\times 10^{24}$ \\ 
&  &  &  &  &  &  &  & (b) & 5.809$\times 10^{24}$ \\ 
&  &  &  &  & \cite{hir94} & QRPA & 0.0875 & (a) & 3.001$\times 10^{22}$ \\ 
&  &  &  &  &  &  &  & (b) & 7.587$\times 10^{22}$ \\ \cline{2-10}
& $ECEC$ & \cite{gav98} & $>$1.1$\times 10^{17}$ &  & * & PHFB & 0.0525 & (a)
& 7.123$\times 10^{21}$ \\ 
&  &  &  &  &  &  &  & (b) & 1.801$\times 10^{22}$ \\ 
&  &  &  &  & \cite{rum98} & SU(4)$_{\sigma \tau }$ & 0.0528 & (a) & 7.032$%
\times 10^{21}$ \\ 
&  &  &  &  &  &  &  & (b) & 1.778$\times 10^{22}$ \\ 
&  &  &  &  & \cite{aun96} & MCM & 0.0071 & (a) & 3.889$\times 10^{23}$ \\ 
&  &  &  &  &  &  &  & (b) & 9.833$\times 10^{23}$ \\ 
&  &  &  &  & \cite{hir94} & QRPA & 0.0822 & (a) & 2.901$\times 10^{21}$ \\ 
&  &  &  &  &  &  &  & (b) & 7.336$\times 10^{21}$ \\ \hline
$^{126}$Xe & $ECEC$ &  &  &  & * & PHFB & 0.0487 & (a) & 5.682$\times
10^{24} $ \\ 
&  &  &  &  &  &  &  & (b) & 1.437$\times 10^{25}$ \\ \hline
$^{130}$Ba & $\beta ^{+}\beta ^{+}$ & \cite{mes01} & (2.16$\pm 0.52)\times
10^{21\dagger}$ &  & * & PHFB & 0.0415 & (a) & 4.797$\times 10^{29}$ \\ 
&  & \cite{bar96a} & $>$4.0$\times 10^{21**}$ &  &  &  &  & (b) & 1.213$%
\times 10^{30}$ \\ 
&  &  &  &  & \cite{hir94} & QRPA & 0.0697 & (a) & 1.700$\times 10^{29}$ \\ 
&  &  &  &  &  &  &  & (b) & 4.298$\times 10^{29}$ \\ 
&  &  &  &  & \cite{sta91} & QRPA &  &  & 1.370$\times 10^{29}$ \\ 
\cline{2-10}
& $\beta ^{+}EC$ & \cite{mes01} & (2.16$\pm 0.52)\times 10^{21\dagger}$ &  & 
* & PHFB & 0.0415 & (a) & 4.188$\times 10^{23}$ \\ 
&  & \cite{bar96a} & $>$4.0$\times 10^{21**}$ &  &  &  &  & (b) & 1.059$%
\times 10^{24}$ \\ 
&  &  &  &  & \cite{rum98} & SU(4)$_{\sigma \tau }$ & 0.0568 & (a) & 2.235$%
\times 10^{23}$ \\ 
&  &  &  &  &  &  &  & (b) & 5.651$\times 10^{23}$ \\ 
&  &  &  &  & \cite{aun96} & MCM & 0.0152 & (a) & 3.121$\times 10^{24}$ \\ 
&  &  &  &  &  &  &  & (b) & 7.890$\times 10^{24}$ \\ 
&  &  &  &  & \cite{hir94} & QRPA & 0.0849 & (a) & 1.000$\times 10^{23}$ \\ 
&  &  &  &  &  &  &  & (b) & 2.529$\times 10^{23}$ \\ \cline{2-10}
& $ECEC$ & \cite{mes01} & (2.16$\pm 0.52)\times 10^{21\dagger}$ &  & * & PHFB
& 0.0415 & (a) & 1.405$\times 10^{22}$ \\ 
&  & \cite{bar96a} & $>$4.0$\times 10^{21**}$ &  &  &  &  & (b) & 3.553$%
\times 10^{22}$ \\ 
&  &  &  &  & \cite{rum98} & SU(4)$_{\sigma \tau }$ & 0.0568 & (a) & 7.498$%
\times 10^{21}$ \\ 
&  &  &  &  &  &  &  & (b) & 1.896$\times 10^{22}$ \\ 
&  &  &  &  & \cite{aun96} & MCM & 0.0072 & (a) & 4.666$\times 10^{23}$ \\ 
&  &  &  &  &  &  &  & (b) & 1.180$\times 10^{24}$ \\ 
&  &  &  &  & \cite{hir94} & QRPA & 0.0759 & (a) & 4.199$\times 10^{21}$ \\ 
&  &  &  &  &  &  &  & (b) & 1.062$\times 10^{22}$ \\ \hline
$^{132}$Ba & $ECEC$ & \cite{mes01} & (1.3$\pm 0.9)\times 10^{21\dagger}$ & 
& * & PHFB & 0.0522 & (a) & 5.474$\times 10^{24}$ \\ 
&  & \cite{bar96a} & $>$3.0$\times 10^{20**}$ &  &  &  &  & (b) & 1.384$%
\times 10^{25}$ \\ \hline\hline
\end{tabular}

\noindent * denotes present work, ** denotes ($0\nu +2\nu )$ decay mode, $%
^\dagger $ denotes ($0\nu +2\nu +0\nu M)$ decay mode

\noindent

\smallskip \noindent \noindent \noindent \textbf{Table 5:} Effect of the
variation in $\zeta _{qq}$ on $\left\langle Q_{0}^{2}\right\rangle ,$ $\beta
_{2}$ and M$_{2\nu }$ for $^{124.126,128,130}$Te, $^{124,126,128,,130,132}$%
Xe, and $^{130,132}$Ba nuclei.

\begin{tabular}{c}
\end{tabular}

\noindent \noindent \noindent 
\begin{tabular}{p{0.22in}p{0.2in}llllllllllllll}
\hline\hline
{\small Nuclei} & $\zeta _{qq}$ & {\small 0.00} & {\small 0.20} & {\small %
0.40} & {\small 0.60} & {\small 0.80} & {\small 0.90} & {\small 0.95} & 
{\small 1.00} & {\small 1.05} & {\small 1.10} & {\small 1.20} & {\small 1.30}
& {\small 1.40} & {\small 1.50} \\ \hline
&  &  &  &  &  &  &  &  &  &  &  &  &  &  &  \\ 
$^{124}${\small Xe} & $\left\langle Q_{0}^{2}\right\rangle $ & {\small 0.00}
& {\small 0.310} & {\small 1.144} & {\small 36.492} & {\small 62.41} & 
{\small 69.16}$^{*}$ & {\small 71.03} & {\small 73.36} & {\small 76.17} & 
{\small 79.81} & {\small 85.96} & {\small 89.47}$^{\dagger }$ & {\small 90.66%
} & {\small 91.87} \\ 
& $\beta _{2}$ & {\small 0.00} & {\small 0.093} & {\small 0.094} & {\small %
0.135} & {\small 0.190} & {\small 0.202}$^{*}$ & {\small 0.205} & {\small %
0.210} & {\small 0.216} & {\small 0.225} & {\small 0.245} & {\small 0.259}$%
^{\dagger }$ & {\small 0.261} & {\small 0.262} \\ 
$^{124}${\small Te} & $\left\langle Q_{0}^{2}\right\rangle $ & {\small 0.00}
& {\small 0.124} & {\small 0.506} & {\small 33.151} & {\small 47.27} & 
{\small 52.98}$^{*}$ & {\small 55.03} & {\small 57.01} & {\small 58.91} & 
{\small 60.69} & {\small 63.98} & {\small 67.41}$^{\dagger }$ & {\small 71.15%
} & {\small 76.14} \\ 
& $\beta _{2}$ & {\small 0.00} & {\small 0.077} & {\small 0.079} & {\small %
0.123} & {\small 0.148} & {\small 0.157}$^{*}$ & {\small 0.161} & {\small %
0.164} & {\small 0.167} & {\small 0.170} & {\small 0.176} & {\small 0.181}$%
^{\dagger }$ & {\small 0.188} & {\small 0.197} \\ 
& $M_{2\nu }$ & {\small 0.189} & {\small 0.195} & {\small 0.193} & {\small %
0.116} & {\small 0.075} & {\small 0.060}$^{*}$ & {\small 0.057} & {\small %
0.052} & {\small 0.048} & {\small 0.041} & {\small 0.025} & {\small 0.023}$%
^{\dagger }$ & {\small 0.028} & {\small 0.033} \\ 
&  &  &  &  &  &  &  &  &  &  &  &  &  &  &  \\ 
$^{126}${\small Xe} & $\left\langle Q_{0}^{2}\right\rangle $ & {\small 0.0}
& {\small 0.122} & {\small 0.473} & {\small 33.73} & {\small 55.53} & 
{\small 61.80} & {\small 64.97} & {\small 67.34} & {\small 69.26} & {\small %
70.53} & {\small 77.56} & {\small 92.05} & {\small 94.74} & {\small 95.24}
\\ 
& $\beta _{2}$ & {\small 0.00} & {\small 0.088} & {\small 0.092} & {\small %
0.128} & {\small 0.174} & {\small 0.187} & {\small 0.194} & {\small 0.199} & 
{\small 0.203} & {\small 0.204} & {\small 0.218} & {\small 0.262} & {\small %
0.266} & {\small 0.267} \\ 
$^{126}${\small Te} & $\left\langle Q_{0}^{2}\right\rangle $ & {\small 0.00}
& {\small 0.058} & {\small 0.541} & {\small 29.31} & {\small 41.23} & 
{\small 46.00} & {\small 48.04} & {\small 49.83} & {\small 51.45} & {\small %
52.98} & {\small 55.32} & {\small 58.14} & {\small 61.57} & {\small 66.99}
\\ 
& $\beta _{2}$ & {\small 0.00} & {\small 0.073} & {\small 0.079} & {\small %
0.114} & {\small 0.136} & {\small 0.145} & {\small 0.148} & {\small 0.151} & 
{\small 0.153} & {\small 0.156} & {\small 0.160} & {\small 0.164} & {\small %
0.170} & {\small 0.179} \\ 
& $M_{2\nu }$ & {\small 0.170} & {\small 0.197} & {\small 0.169} & {\small %
0.104} & {\small 0.077} & {\small 0.063} & {\small 0.055} & {\small 0.049} & 
{\small 0.047} & {\small 0.048} & {\small 0.039} & {\small 0.007} & {\small %
0.002} & {\small 0.004} \\ 
&  &  &  &  &  &  &  &  &  &  &  &  &  &  &  \\ 
$^{128}${\small Te} & $\left\langle Q_{0}^{2}\right\rangle $ & {\small 0.00}
& {\small 0.019} & {\small 0.419} & {\small 2.662} & {\small 34.01} & 
{\small 38.08} & {\small 39.94} & {\small 41.62} & {\small 43.14} & {\small %
44.61} & {\small 47.13} & {\small 49.34} & {\small 50.81}$^{\ddagger }$ & 
{\small 52.98} \\ 
& $\beta _{2}$ & {\small 0.00} & {\small 0.058} & {\small 0.079} & {\small %
0.079} & {\small 0.122} & {\small 0.129} & {\small 0.133} & {\small 0.136} & 
{\small 0.138} & {\small 0.141} & {\small 0.145} & {\small 0.148} & {\small %
0.150}$^{\ddagger }$ & {\small 0.154} \\ 
$^{128}${\small Xe} & $\left\langle Q_{0}^{2}\right\rangle $ & {\small 0.00}
& {\small 0.329} & {\small 0.964} & {\small 42.52} & {\small 57.17} & 
{\small 61.30} & {\small 62.86} & {\small 64.54} & {\small 67.90} & {\small %
72.94} & {\small 90.48} & {\small 96.90} & {\small 99.83}$^{\ddagger }$ & 
{\small 100.05} \\ 
& $\beta _{2}$ & {\small 0.00} & {\small 0.091} & {\small 0.092} & {\small %
0.146} & {\small 0.178} & {\small 0.186} & {\small 0.189} & {\small 0.192} & 
{\small 0.199} & {\small 0.209} & {\small 0.257} & {\small 0.269} & {\small %
0.275}$^{\ddagger }$ & {\small 0.275} \\ 
& $M_{2\nu }$ & {\small 0.142} & {\small 0.138} & {\small 0.139} & {\small %
0.081} & {\small 0.042} & {\small 0.036} & {\small 0.033} & {\small 0.033} & 
{\small 0.029} & {\small 0.026} & {\small 0.003} & {\small 0.0004} & {\small %
0.00002}$^{\ddagger }$ & {\small 0.00003} \\ 
&  &  &  &  &  &  &  &  &  &  &  &  &  &  &  \\ 
$^{130}${\small Te} & $\left\langle Q_{0}^{2}\right\rangle $ & {\small 0.00}
& {\small -0.003} & {\small 0.040} & {\small 1.654} & {\small 0.316} & 
{\small 26.80} & {\small 29.35} & {\small 32.12} & {\small 33.69} & {\small %
35.03} & {\small 37.37} & {\small 39.30} & {\small 41.02} & {\small 42.65}
\\ 
& $\beta _{2}$ & {\small 0.00} & {\small 0.006} & {\small 0.065} & {\small %
0.078} & {\small 0.079} & {\small 0.100} & {\small 0.107} & {\small 0.117} & 
{\small 0.120} & {\small 0.123} & {\small 0.127} & {\small 0.130} & {\small %
0.133} & {\small 0.136} \\ 
$^{130}${\small Xe} & $\left\langle Q_{0}^{2}\right\rangle $ & {\small 0.00}
& {\small 0.189} & {\small 0.631} & {\small 22.26} & {\small 43.56} & 
{\small 48.61} & {\small 50.84} & {\small 52.17} & {\small 54.41} & {\small %
55.37} & {\small 57.92} & {\small 61.13} & {\small 64.45} & {\small 77.79}
\\ 
& $\beta _{2}$ & {\small 0.00} & {\small 0.091} & {\small 0.092} & {\small %
0.103} & {\small 0.147} & {\small 0.159} & {\small 0.164} & {\small 0.166} & 
{\small 0.171} & {\small 0.173} & {\small 0.179} & {\small 0.186} & {\small %
0.190} & {\small 2.155} \\ 
& $M_{2\nu }$ & {\small 0.123} & {\small 0.124} & {\small 0.124} & {\small %
0.116} & {\small 0.061} & {\small 0.057} & {\small 0.049} & {\small 0.042} & 
{\small 0.038} & {\small 0.036} & {\small 0.031} & {\small 0.025} & {\small %
0.025} & {\small 0.012} \\ 
&  &  &  &  &  &  &  &  &  &  &  &  &  &  &  \\ 
$^{130}${\small Ba} & $\left\langle Q_{0}^{2}\right\rangle $ & {\small 0.00}
& {\small 0.145} & {\small 0.617} & {\small 56.36} & {\small 67.57} & 
{\small 71.39} & {\small 73.18} & {\small 79.29} & {\small 93.74} & {\small %
98.74} & {\small 105.5} & {\small 108.6} & {\small 109.4} & {\small 110.5}
\\ 
& $\beta _{2}$ & {\small 0.00} & {\small 0.092} & {\small 0.095} & {\small %
0.190} & {\small 0.212} & {\small 0.220} & {\small 0.223} & {\small 0.234} & 
{\small 0.272} & {\small 0.286} & {\small 0.298} & {\small 0.307} & {\small %
0.310} & {\small 0.314} \\ 
$^{130}${\small Xe} & $\left\langle Q_{0}^{2}\right\rangle $ & {\small 0.00}
& {\small 0.189} & {\small 0.631} & {\small 22.26} & {\small 43.56} & 
{\small 47.96} & {\small 50.84} & {\small 52.17} & {\small 54.41} & {\small %
55.37} & {\small 57.92} & {\small 61.13} & {\small 64.45} & {\small 77.79}
\\ 
& $\beta _{2}$ & {\small 0.00} & {\small 0.091} & {\small 0.092} & {\small %
0.103} & {\small 0.147} & {\small 0.157} & {\small 0.164} & {\small 0.166} & 
{\small 0.171} & {\small 0.173} & {\small 0.179} & {\small 0.186} & {\small %
0.190} & {\small 0.216} \\ 
& $M_{2\nu }$ & {\small 0.191} & {\small 0.188} & {\small 0.190} & {\small %
0.028} & {\small 0.033} & {\small 0.030} & {\small 0.032} & {\small 0.041} & 
{\small 0.009} & {\small 0.002} & {\small 0.0002} & {\small 0.0001} & 
{\small 0.0002} & {\small 0.002} \\ 
&  &  &  &  &  &  &  &  &  &  &  &  &  &  &  \\ 
$^{132}${\small Ba} & $\left\langle Q_{0}^{2}\right\rangle $ & {\small 0.00}
& {\small 0.067} & {\small 0.323} & {\small 0.750} & {\small 1.660} & 
{\small 33.83} & {\small 46.33} & {\small 50.29} & {\small 52.36} & {\small %
54.33} & {\small 57.70} & {\small 60.30} & {\small 62.99} & {\small 64.55}
\\ 
& $\beta _{2}$ & {\small 0.00} & {\small 0.085} & {\small 0.094} & {\small %
0.096} & {\small 0.096} & {\small 0.125} & {\small 0.163} & {\small 0.176} & 
{\small 0.180} & {\small 0.184} & {\small 0.191} & {\small 0.196} & {\small %
0.202} & {\small 0.205} \\ 
$^{132}${\small Xe} & $\left\langle Q_{0}^{2}\right\rangle $ & {\small 0.00}
& {\small 0.009} & {\small 0.103} & {\small 0.495} & {\small 3.437} & 
{\small 32.64} & {\small 35.37} & {\small 37.75} & {\small 39.54} & {\small %
41.12} & {\small 44.70} & {\small 46.38} & {\small 48.92} & {\small 51.33}
\\ 
& $\beta _{2}$ & {\small 0.00} & {\small 0.039} & {\small 0.089} & {\small %
0.092} & {\small 0.092} & {\small 0.121} & {\small 0.127} & {\small 0.133} & 
{\small 0.137} & {\small 0.141} & {\small 0.150} & {\small 0.153} & {\small %
0.158} & {\small 0.163} \\ 
& $M_{2\nu }$ & {\small 0.161} & {\small 0.164} & {\small 0.166} & {\small %
0.167} & {\small 0.170} & {\small 0.142} & {\small 0.086} & {\small 0.052} & 
{\small 0.049} & {\small 0.046} & {\small 0.043} & {\small 0.040} & {\small %
0.037} & {\small 0.034} \\ 
&  &  &  &  &  &  &  &  &  &  &  &  &  &  &  \\ 
$^{150}${\small Nd} & $\left\langle Q_{0}^{2}\right\rangle $ & {\small 0.00}
& {\small 0.070} & {\small 0.170} & {\small 23.55} & {\small 51.13} & 
{\small 64.98} & {\small 77.06} & {\small 83.75} & {\small 86.89} & {\small %
87.76} & {\small 88.40} & {\small 88.93} & {\small 89.37} & {\small 89.73}
\\ 
& $\beta _{2}$ & {\small 0.00} & {\small 0.066} & {\small 0.075} & {\small %
0.096} & {\small 0.163} & {\small 0.214} & {\small 0.257} & {\small 0.276} & 
{\small 0.283} & {\small 0.285} & {\small 0.288} & {\small 0.290} & {\small %
0.291} & {\small 0.293} \\ 
$^{150}${\small Sm} & $\left\langle Q_{0}^{2}\right\rangle $ & {\small 0.00}
& {\small 0.104} & {\small 0.649} & {\small 1.477} & {\small 33.14} & 
{\small 48.66} & {\small 58.45} & {\small 73.41} & {\small 83.53} & {\small %
88.11} & {\small 93.23} & {\small 97.48} & {\small 99.85} & {\small 101.99}
\\ 
& $\beta _{2}$ & {\small 0.00} & {\small 0.067} & {\small 0.081} & {\small %
0.084} & {\small 0.112} & {\small 0.155} & {\small 0.187} & {\small 0.238} & 
{\small 0.268} & {\small 0.281} & {\small 0.297} & {\small 0.309} & {\small %
0.314} & {\small 0.319} \\ 
& $M_{2\nu }$ & {\small 0.196} & {\small 0.199} & {\small 0.201} & {\small %
0.200} & {\small 0.140} & {\small 0.113} & {\small 0.056} & {\small 0.033} & 
{\small 0.028} & {\small 0.027} & {\small 0.029} & {\small 0.026} & {\small %
0.022} & {\small 0.015} \\ \hline\hline
\end{tabular}

\noindent \noindent $^{*}$denotes $\zeta _{qq}=0.91,$ $^{\dagger }$denotes $%
\zeta _{qq}=1.31,$ $^{\ddagger }$denotes $\zeta _{qq}=1.41$.


\begin{thebibliography}{99}
\bibitem{pri52}  H. Primakoff, Phys. Rev. \textbf{85}, 888 (1952); S.P.
Rosen, H. Primakoff, Report Prog. in Phys. \textbf{22}, 121 (1959); S.P.
Rosen, H .Primakoff, \textit{Alpha-beta-gamma ray spectroscopy}, ed. K.
Siegbahn (1965) p.1499; H. Primakoff, S.P. Rosen, Ann. Rev. Nucl. Part. Sci. 
\textbf{31}, 145 (1981).

\bibitem{bri78}  D. Bryman, C. Picciotto, Rev. Mod. Phys. \textbf{50}, 11
(1978).

\bibitem{hax84}  W. C. Haxton, G. J. Stephenson, Jr., Prog. Part. Nucl.
Phys. \textbf{12}, 409 (1984).

\bibitem{doi85}  M. Doi, T. Kotani, E. Takasugi, Prog. Theo. Phys. Suppl. 
\textbf{83}, 1 (1985).

\bibitem{ver86}  J. D. Vergados, Nucl. Phys. B \textbf{218}, 109 (1983),
Phys. Rep. \textbf{133}, 1 (1986); ibid \textbf{361}, 1 (2002).

\bibitem{fae88}  A. Faessler, Prog. Part. Nucl. Phys. \textbf{21}, 183
(1988).

\bibitem{tom91}  T. Tomoda, Rep. Prog. Phys. \textbf{54}, 53 (1991).

\bibitem{boe92}  F. Boehm, P. Vogel, Ann. Rev. Nucl. Part. Sci. \textbf{34},
125 (1984); F. Boehm, P. Vogel, \textit{Physics of Massive Neutrinos}, 2nd
ed. (Cambridge University Press, Cambridge 1992).

\bibitem{zub98}  K. Zuber, Phys. Rep. \textbf{305}, 295 (1998).

\bibitem{fio98}  E. Fiorini, Phys. Rep. \textbf{307}, 309 (1998).

\bibitem{eji00}  H. Ejiri, Phys. Rep. \textbf{338}, 265 (2000).

\bibitem{ell02}  S. R. Elliott and J. Engel, J. Phys. G: Nucl. Part. Phys. 
\textbf{30} (2004) R183.

\bibitem{tre95}  V.I. Tretyak, Y.G. Zdesenko, At. Data Nucl. Data Tables 
\textbf{61}, 43 (1995); ibid \textbf{80}, 83 (2002).

\bibitem{kla04}  H.V. Klapdor-Kleingrothaus, et al., Mod. Phys. Lett. 
\textbf{16} (2001) 2409; H.V. Klapdor-Kleingrothaus, et al., Phys. Lett. 
\textbf{B 586} (2004) 198.

\bibitem{doi92}  M. Doi, T. Kotani, Prog. Theor. Phy. \textbf{87}, 1207
(1992).

\bibitem{doi93}  M. Doi, T. Kotani, Prog. Theor. Phy. \textbf{89}, 139
(1993).

\bibitem{moe94}  M.K. Moe, P. Vogel, Ann. Rev. Nucl. Part. Sci. \textbf{44},
247 (1994).

\bibitem{suh98}  J. Suhonen, O. Civitarese, Phys. Rep. \textbf{300}, 123
(1998).

\bibitem{fae98}  A. Faessler, F. Simkovic, hep-ph/9901215; J. Phys. G 
\textbf{24,} 2139 (1998).

\bibitem{kla01}  H.V. Klapdor-Kleingrothaus, \textit{Sixty years of Double
Beta Decay, }World Scientific, Singapore (2001).

\bibitem{bar96}  A.S. Barabash, V.I. Umatov, R. Gurriaran, F. Hubert, Ph.
Hubert, M. Aunola, J. Suhonen, Nucl. Phys. A \textbf{604}, 115 (1996).

\bibitem{civ00}  O. Civitarese, J. Suhonen, Phys. Lett. B \textbf{482}, 368
(2000).

\bibitem{suh01}  J. Suhonen, O. Civitarese, Phys. Lett. B\textbf{\ 497}, 221
(2001).

\bibitem{gri92}  A. Griffiths, P. Vogel, Phys. Rev. C \textbf{46}, 181
(1992).

\bibitem{suh94}  J. Suhonen, O. Civitarese, Phys. Rec. C \textbf{49}, 3055
(1994).

\bibitem{aue93}  N. Auerbach, D.C. Zheng, L. Zamick, B.A. Brown, Phys. Lett.B%
\textbf{\ 304} (1993) 17; N. Auerbach, G.F. Bertsch, B.A. Brown, L. Zhao,
Nucl. Phys. A\textbf{\ 56}, 190 (1993).

\bibitem{tro96}  D. Troltenier, J.P. Draayer, J.G. Hirsch, Nucl. Phys. A 
\textbf{\ 601}, 89 (1996).

\bibitem{fri95}  F. Frisk, I. Hamamoto, X.Z. Zhang, Phys. Rev. C\textbf{\ 52}%
, 2468 (1995).

\bibitem{sar98}  P. Sarriguren, E. Moya de Guerra, A. Escuderos, A.C.
Carrizo, Nucl. Phys. A\textbf{\ 635}, 55 (1998); P. Sarriguren, E. Moya de
Guerra, A. Escuderos, Nucl. Phys. A\textbf{\ 658} (1999) 13; \textit{ibid}
Nucl. Phys. A\textbf{\ 691}, 631 (2001).

\bibitem{nach04}  E. N\'{a}cher \textit{et al.}, Phys. Rev. Lett. \textbf{92}%
, 232501(2004).

\bibitem{pac04}  L. Pacearescu, A. Faessler, F. Simkovic, Phys. At. Nucl. 
\textbf{67}, 1210 (2004).

\bibitem{alv04}  R. \'{A}lvarez-Rodr\'{i}guez, P. Sarriguren, E. Moya de
Guerra, L. Pacearescu, A. Faessler, F. Simkovic, Phys. Rev. C \textbf{70},
064309 (2004).

\bibitem{cha05}  R. Chandra, J. Singh, P.K. Rath, P.K. Raina, J.G. Hirsch,
Eur. Phys. J. A\textbf{\ 23}, 223 (2005)

\bibitem{rai06}  P.K. Raina, A. Shukla, S. Singh, P.K. Rath, J.G. Hirsch,
Eur. Phys. J. A\textbf{\ 28}, 27 (2006); A. Shukla, P.K. Raina, R. Chandra,
P.K. Rath, J.G. Hirsch, Eur. Phys. J. A\textbf{\ 23}, 235 (2005).

\bibitem{shuk07}  A. Shukla, P.K. Raina, P.K. Rath, J.Phys. G: Nucl. Part.
Phys. \textbf{34}, 549 (2007).

\bibitem{bar68}  M. Baranger, K. Kumar, Nucl. Phys. A\textbf{\ 110}, 490
(1968).

\bibitem{dix02}  B.M. Dixit, P.K. Rath, P.K. Raina, Phys. Rev. C\textbf{\ } 
\textbf{65}, 034311 (2002), Phys. Rev. C \textbf{67}, 059901(E) (2003).

\bibitem{civ93}  O. Civitarese, J. Suhonen, Phys. Rev. C\textbf{\ 47}, 2410
(1993).

\bibitem{wu91}  X. R. Wu, A. Staudt, H. V. Klapdor-Kleingrothaus, Chen-rui
Ching and Tso-hsiu Ho, Phys. Lett. B \textbf{272}, 169 (1991); X. R. Wu, A.
Staudt, T. T. S. Kuo and H. V. Klapdor-Kleingrothaus, Phys. Lett. B \textbf{%
276}, 274 (1992).

\bibitem{eng92}  J. Engel, W. C. Haxton and P. Vogel, Phys. Rev. C \textbf{46%
}, R2153 (1992)

\bibitem{vog86}  P. Vogel, M.R. Zirnbauer, Phys. Rev. Lett. \textbf{57},
3148 (1986).

\bibitem{cas94}  O. Casta\~{n}os, J.G. Hirsch, O. Civitarese, P.O. Hess,
Nucl. Phys. A \textbf{571}, 276 (1994).

\bibitem{hir95}  J.G. Hirsch, O. Casta\~{n}os, P.O. Hess, O. Civitarese,
Phys. Rev. C\textbf{\ 51}, 2252 (1995).

\bibitem{cer99}  V.E. Ceron, J.G. Hirsch, Phys. Lett.B \textbf{471}, 1
(1999).

\bibitem{oni66}  N. Onishi, S. Yoshida, Nucl. Phys. A\textbf{\ 260}, 226
(1966).

\bibitem{boh98}  A. Bohr, B.R. Mottelson, Nuclear Structure Vol. I (World
Scientific, Singapore, 1998).

\bibitem{ran97}  Rani Devi, S.P. Sarswat, Arun Bharti, S.K. Khosa, Phys.
Rev. C \textbf{55}, 2433 (1997).

\bibitem{ari81}  A. Arima, Nucl. Phys. A\textbf{\ 354}, 19 (1981).

\bibitem{sak84}  M. Sakai At. Data, Nucl. Data Tables \textbf{31} ,
400(1984).

\bibitem{kho82}  S.K. Khosa, P.N. Tripathi, S.K. Sharma, Phys. Lett. B 
\textbf{119}, 257 (1982); P.N. Tripathi, S.K. Sharma, S.K. Khosa, Phys. Rev.
C \textbf{29}, 1951 (1984); S.K. Sharma, P.N. Tripathi, S. K. Khosa, Phys.
Rev. C \textbf{38}, 2935 (1988).

\bibitem{ram87}  S. Raman, C.H. Malarkey, W.T. Milner, C.W. Nestor, Jr.,
P.H. Stelson, At. Data Nucl. Data Tables \textbf{36}, 1 (1987).

\bibitem{rag89}  P. Raghavan, At. Data Nucl. Data Table \textbf{42}, 189
(1989).

\bibitem{ram01}  S. Raman, C. W. Nestor, Jr., P. Tikkanen, At. Data Nucl.
Data Tables \textbf{78} (2001).

\bibitem{tak96}  N. Takaoka \textit{et al.}, Phys. Rev. C \textbf{53}, 1557
(1996).

\bibitem{ber92}  T. Bernatovicz, J. Brannon, R. Brazzle, R. Cowsik, C.
Hohenberg, F. Podosek, Phys. Rev. Lett. \textbf{69}, 2341 (1992); Phys. Rev.
C \textbf{47}, 806 (1993).

\bibitem{bar02}  A.S. Barabash, Czech. J. Phys. \textbf{52,} 567 (2002);
nucl-ex/0203001 (2002).

\bibitem{lin88}  J.W. Lin \textit{et al.}, Nucl. Phys. A \textbf{481}, 484
(1988).

\bibitem{man86}  O.K. Manuel, Proc. Int. Symp. on Nuclear Beta Decays,
Neutrino, Osaka, Japan, June 1986 (World Sci., 1986), pg. 71.

\bibitem{rum98}  O.A. Rumyantsev, M.H. Urin, Phys. Lett. B \textbf{443}, 51
(1998).

\bibitem{sem00}  S.V. Semenov, F. Simkovic, V.V. Khruschev, P. Domin, Phys.
At. Nucl. \textbf{63}, 1196 (2000).

\bibitem{aun96}  M. Aunola, J. Suhonen, Nucl. Phys. A \textbf{602}, 133
(1996).

\bibitem{civ98}  O. Civitarese, J. Suhonen, Phys. Rev. C \textbf{58}, 1535
(1998).

\bibitem{man91}  O.K. Manuel, J. Phys. G\textbf{17}, S221 (1991).

\bibitem{eng88}  J. Engel, P. Vogel, M.R. Zirnbauer, Phys. Rev. C \textbf{37}%
, 731 (1988).

\bibitem{sto94}  S. Stoica, Phys. Rev. C \textbf{49}, 2240 (1994).

\bibitem{sta90}  A. Staudt, K. Muto, H.V. Klapdor, Euro. Phys. Lett. \textbf{%
13}, 31 (1990).

\bibitem{hir94a}  M. Hirsch, X.R. Wu, H.V. Klapdor-Kleingrothaus, Ching
Cheng-rui, Ho Tso-hsiu, Phys. Rep. \textbf{242}, 403 (1994).

\bibitem{sch85}  O. Scholten, Z.R. Yu, Phys. Lett. B \textbf{161}, 13 (1985).

\bibitem{arn03}  C. Arnaboldi \textit{et al}., Phys. Lett. B \textbf{557},
167 (2003)

\bibitem{cau99}  E. Caurier, F. Nowacki, A. Poves, J. Retamosa, Nucl. Phys.
A \textbf{654}, 973 (1999).

\bibitem{toi97}  J. Toivanen, J. Suhonen, Phys. Rev. C \textbf{55}, 2314
(1997).

\bibitem{sil97}  A. De Silva, M.K. Moe, M.A. Nelson, M.A. Vient, Phys. Rev.
C \textbf{56}, 2451 (1997).

\bibitem{lal05}  D. Lalanne, hep-ex/0509005, Dans High Energy Physics ICHEP
2004 (in 2 volumes), World Scientific (Ed.) (2005).

\bibitem{art95}  V. Artemiev \textit{et al.}, Phys. Lett. B \textbf{345},
564 (1995).

\bibitem{kir86}  T. Kirsten \textit{et al.}, Proc. Int. Symp.``Nuclear Beta
Decay, Neutrino (Osaka 86)'', World Sci., Singapore, 81 (1986).

\bibitem{all00}  A. Allessandrello \textit{et al.}, Phys. Lett. B \textbf{486%
}, 13 (2000).

\bibitem{bel87}  E. Bellotti, C. Cattadori, O. Cremonesi, E. Fiorini, C.
Liguori, A. Pullia, P.P. Sverzellati, L. Zanotti, Euro. Phys. Lett. \textbf{3%
}, 889 (1987).

\bibitem{zde80}  Yu. G. Zdesenko, Sov. J. Part. Nucl. \textbf{11(6)}, 542
(1980).

\bibitem{rum95}  O.A. Rumyantsev, M.H. Urin, JETP Lett. \textbf{61}, 361
(1995).

\bibitem{vas90}  S.I. Vasilev \textit{et al.}, JETP Lett. \textbf{51}, 622
(1990); \textbf{58}, 178 (1993).

\bibitem{art93}  V. Artemiev \textit{et al.}, JETP Lett. \textbf{58}, 262
(1993).

\bibitem{ell93}  S.R. Elliott, M.K. Moe, M.A. Nelson, M.A. Vient, Nucl.
Phys. B (Proc. Suppl.) \textbf{31}, 68 (1993).

\bibitem{kli86}  A.A. Klimenko, A.A. Pomansky, A.A. Smolnikov, Nucl.
Instrum. Meth. B \textbf{17}, 445 (1986).

\bibitem{bol97}  A. Bolozdynya \textit{et. al.}, IEEE Nuclear Science
Symposium, Medical Imaging Conference, \textbf{2}, 697 (1996); ibid IEEE
Transaction on Nuclear Science, \textbf{44}, 3 (1997).

\bibitem{gav98}  Ju.M. Gavriljuk, V.V. Kuzminov, N.Yu. Osetrova, S.S.
Ratkevich\textit{,} G.V. Volchenko, Phys. At. Nucl. \textbf{61}, 1287 (1998).

\bibitem{bar89}  A.S. Barabash, V.V. Kuzminov, V.M. Lobashev, V.M. Novikov,
B.M. Ovchinnikov, A.A. Pomansky, Phys. Lett. B \textbf{223}, 273 (1989).

\bibitem{hir94}  M. Hirsch, K. Muto, T. Oda, H.V. Klapdor- Kleingrothaus, Z.
Phys. A \textbf{347}, 151 (1994).

\bibitem{sta91}  A. Staudt, K. Muto, H.V. Klapdor-Kleingrothaus, Phys. Lett.
B \textbf{268}, 312 (1991).

\bibitem{mes01}  A.P. Meshik, C.M. Hohenberg, O.V. Pravdivtseva, Ya. S.
Kapusta, Phys. Rev. C \textbf{64}, 035205 (2001).

\bibitem{bar96a}  A. S. Barabash, R.R. Saakyan, Phys. At. Nucl. \textbf{59},
179 (1996).

\pagebreak 
\end{thebibliography}
\end{document}